\documentclass[journal,onecolumn,draftclsnofoot]{IEEEtran}

\let\svtikzpicture\tikzpicture
\def\tikzpicture{\noindent\svtikzpicture}
\usepackage[caption=false]{subfig}
\usepackage[thinc]{esdiff}
\usepackage{commath,amsmath,amssymb,amsfonts}
\usepackage{mathtools}
\usepackage{amsthm}
\usepackage{algorithmic}
\usepackage{pgfplots} 
\pgfplotsset{compat=1.15}
\usepackage{graphicx}
\usepackage{pgfgantt}
\usepackage{pdflscape}
\usepackage{pst-plot}
\usepackage{xfrac}
\usepackage{colortbl}
\usepackage{cancel}
\usepgfplotslibrary{fillbetween}
\usepackage{amssymb,bm}
\usepackage{float}
\usepackage{amsmath}
\usepackage[draft=false]{hyperref}
\usepackage{multirow}
\usepackage{xcolor}
\usepackage{mathrsfs}
\usepackage{bbm}

\usepackage{cite} 

\usepackage{comment}

\def \B{\mathcal{B}}

\def \D{\mathcal{D}}
\def \E{\mathcal{E}}

\def \H{\mathcal{H}}

\def \M{\mathcal{M}}

\def \Q{\mathcal{Q}}
\def \R{\mathcal{R}}
\def \S{\mathcal{S}}

\def \U{\mathcal{U}}

\def \W{\mathcal{W}}
\def \X{\mathcal{X}}
\def \Y{\mathcal{Y}}
\def \Z{\mathcal{Z}}

\def \fu{\mathbf{u}}

\def \fx{\mathbf{x}}
\def \fy{\mathbf{y}}

\def \fY{\mathbf{Y}}

\def \f0{\mathbf{0}}

\definecolor{blau_1a}{RGB}{93,133,195}
\definecolor{blau_2a}{RGB}{0,156,218}
\definecolor{gruen_3a}{RGB}{80,182,149}
\definecolor{gruen_4a}{RGB}{175,204,80}
\definecolor{gruen_5a}{RGB}{221,223,72}
\definecolor{orange_6a}{RGB}{255,224,92}
\definecolor{orange_7a}{RGB}{248,186,60}
\definecolor{rot_8a}{RGB}{238,122,52}
\definecolor{rot_9a}{RGB}{233,80,62}
\definecolor{lila_10a}{RGB}{201,48,142}
\definecolor{lila_11a}{RGB}{128,69,151}

\definecolor{blau_1b}{RGB}{0,90,169}
\definecolor{blau_2b}{RGB}{0,131,204}
\definecolor{gruen_3b}{RGB}{0,157,129}
\definecolor{gruen_4b}{RGB}{153,192,0}
\definecolor{gruen_5b}{RGB}{201,212,0}
\definecolor{orange_6b}{RGB}{253,202,0}
\definecolor{orange_7b}{RGB}{245,163,0}
\definecolor{rot_8b}{RGB}{236,101,0}
\definecolor{rot_9b}{RGB}{230,0,26}
\definecolor{lila_10b}{RGB}{166,0,132}
\definecolor{lila_11b}{RGB}{114,16,133}

\definecolor{mycolor1}{rgb}{0.0, 0.18, 0.39}
\definecolor{mycolor2}{RGB}{87,108,67}
\definecolor{mycolor3}{RGB}{8,133,161}
\definecolor{mycolor4}{RGB}{80,91,161}
\definecolor{mycolor5}{RGB}{98,122,157}
\definecolor{mycolor6}{RGB}{255,163,67}
\definecolor{mycolor7}{RGB}{152,205,225}
\definecolor{mycolor8}{RGB}{242,204,48}
\definecolor{mycolor9}{rgb}{0,.5,0}
\definecolor{mycolor10}{rgb}{.59,.44,.09}
\definecolor{mycolor11}{RGB}{231,199,31} % Yellow
\definecolor{mycolor12}{RGB}{8,133,161} % Cyan
\definecolor{mycolor13}{RGB}{157,188,64} % Yellow Green
\definecolor{mycolor14}{RGB}{194,150,130} % Light Skin
\definecolor{mycolor15}{RGB}{98,122,157} % Blue Sky
\definecolor{mycolor16}{RGB}{160,160,160} % Neutral
\definecolor{mycolor17}{RGB}{115,82,68} % Dark Skin
\definecolor{mycolor18}{RGB}{94,60,108} % Purple
\definecolor{mycolor19}{RGB}{115,82,68} % Dark Skin
\definecolor{mycolor20}{RGB}{255,183,30} % Dark Gold

\usepackage{accents}

\theoremstyle{remark} \newtheorem{theorem}{Theorem}
\theoremstyle{remark} \newtheorem{lemma}[theorem]{Lemma}
\theoremstyle{remark} 
\theoremstyle{remark} 
\theoremstyle{remark} \newtheorem{definition}{Definition}
\theoremstyle{remark} 
\theoremstyle{remark} 

\usepackage{amssymb}

\usepackage{autobreak}
\allowdisplaybreaks

\usepackage{tikz}
\usetikzlibrary{arrows.meta}

\usepackage{blochsphere}

\newcommand\pgfmathsinandcos[3]{%
  \pgfmathsetmacro#1{sin(#3)}%
  \pgfmathsetmacro#2{cos(#3)}%
}
\newcommand\LongitudePlane[3][current plane]{%
  \pgfmathsinandcos\sinEl\cosEl{#2} % elevation
  \pgfmathsinandcos\sint\cost{#3} % azimuth
  \tikzset{#1/.style={cm={\cost,\sint*\sinEl,0,\cosEl,(0,0)}}}
}
\newcommand\LatitudePlane[3][current plane]{%
  \pgfmathsinandcos\sinEl\cosEl{#2} % elevation
  \pgfmathsinandcos\sint\cost{#3} % latitude
  \pgfmathsetmacro\yshift{\cosEl*\sint}
  \tikzset{#1/.style={cm={\cost,0,0,\cost*\sinEl,(0,\yshift)}}} %
}
\newcommand\DrawLongitudeCircle[2][1]{
  \LongitudePlane{\angEl}{#2}
  \tikzset{current plane/.prefix style={scale=#1}}
  \pgfmathsetmacro\angVis{atan(sin(#2)*cos(\angEl)/sin(\angEl))} %
  \draw[current plane] (\angVis:1) arc (\angVis:\angVis+180:1);
  \draw[current plane,dashed] (\angVis-180:1) arc (\angVis-180:\angVis:1);
}
\newcommand\DrawLatitudeCircle[2][1]{
  \LatitudePlane{\angEl}{#2}
  \tikzset{current plane/.prefix style={scale=#1}}
  \pgfmathsetmacro\sinVis{sin(#2)/cos(#2)*sin(\angEl)/cos(\angEl)}
  \pgfmathsetmacro\angVis{asin(min(1,max(\sinVis,-1)))}
  \draw[current plane] (\angVis:1) arc (\angVis:-\angVis-180:1);
  \draw[current plane,dashed] (180-\angVis:1) arc (180-\angVis:\angVis:1);
}

\usepackage{multicol, blindtext}

\usepackage[]{pgfplots}
\usepackage{pgfplotstable}
\usepgfplotslibrary{dateplot}
\pgfplotsset{/pgf/number format/use comma,compat=newest}

\usepackage{amsbsy}
\usepackage{bm}
\usepackage{fixmath}

\usepackage{silence}
\usepackage{ctable}

\usepackage{pst-node,pst-plot}

\usetikzlibrary{shapes.geometric}

\usepackage{mathdots}

\usetikzlibrary{math}

\usepackage{tgpagella}
\usepackage[T1]{fontenc}
\usepackage{mdframed}

\allowdisplaybreaks

\begin{document}
% ---put - uaq - rust
\title{\rule{\linewidth}{1pt} \\ \fontfamily{put} \selectfont Deterministic Identification for Molecular Communications over the Poisson Channel \rule{\linewidth}{1pt}}
% ---\line(1,0){400}
\author{Mohammad J. Salariseddigh, Uzi Pereg, Holger Boche, Christian Deppe, \\ Vahid Jamali, and Robert Schober \vspace{-1.4cm}
\thanks{This paper was presented in part at the IEEE Global Communications Conference (GC 2021) \cite{Salariseddigh_GC_IEEE}.}
\thanks{M. J. Salariseddigh, U. Pereg, and C. Deppe are with the Institute for Communications Engineering, Technical University of Munich, Germany (e-mail: \{mjss,~uzi.pereg,~christian.deppe\}@tum.de). H. Boche is with the Chair of Theoretical Information Technology, Technical University of Munich, Germany (e-mail: boche@tum.de). V. Jamali is with the Department of Electrical and Computer Engineering, Princeton University, USA (e-mail: jamali@princeton.edu). R. Schober is with the Institute for Digital Communications, Friedrich-Alexander University Erlangen-N\"urnberg (FAU), Germany (e-mail: robert.schober@fau.de).} 
% \thanks{H. Boche is with the Chair of Theoretical Information Technology, Technical University of Munich (e-mail: boche@tum.de).}
% \thanks{V. Jamali is with the Department of Electrical and Computer Engineering, Princeton University, Princeton, NJ 08544 USA (e-mail: jamali@princeton.edu).}
% \thanks{R. Schober is with the Institute for Digital Communications, Friedrich-Alexander University Erlangen-N\"urnberg (FAU), Erlangen, Germany (e-mail: robert.schober@fau.de).}
}
% ---
\maketitle
% ---
\begin{mdframed}[linewidth=1.5,linecolor=orange_6b, topline=false,rightline=false,bottomline=false]
\textbf{\textcolor{cyan}{Abstract}} \textit{Various applications of molecular communications (MC) are event-triggered, and, as a consequence, the prevalent Shannon capacity may not be the right measure for performance assessment. Thus, in this paper, we motivate and establish the identification capacity as an alternative metric. In particular, we study deterministic identification (DI) for the discrete-time Poisson channel (DTPC), subject to an average and a peak power constraint, which serves as a model for MC systems employing molecule counting receivers. It is established that the codebook size for this channel scales as $2^{(n\log n)R}$, where $n$ and $R$ are the codeword length and coding rate, respectively. Lower and upper bounds on the DI capacity of the DTPC are developed. The obtained large capacity of the DI channel sheds light on the performance of natural DI systems such as natural olfaction, which are known for their extremely large chemical discriminatory power in biology. Furthermore, numerical simulations for the empirical miss-identification and false identification error rates are provided for finite length codes. This allows us to quantify the scale of error reduction in terms of the codeword length.}
\end{mdframed}

% \begin{abstract}
% \rule{\linewidth}{1pt}
% \line(3,0){50}
% \vrule height 50pt
% Various applications of molecular communications (MC) are event-triggered, and, as a consequence, the prevalent Shannon capacity may not be the right measure for performance assessment. Thus, in this paper, we motivate and establish the identification capacity as an alternative metric. In particular, we study deterministic identification (DI) for the discrete-time Poisson channel (DTPC), subject to an average and a peak power constraint, which serves as a model for MC systems employing molecule counting receivers. It is established that the codebook size for this channel scales as $2^{(n\log n)R}$, where $n$ and $R$ are the codeword length and coding rate, respectively. Lower and upper bounds on the DI capacity of the DTPC are developed. The obtained large capacity of the DI channel sheds light on the performance of natural DI systems such as natural olfaction, which are known for their extremely large chemical discriminatory power in biology. Furthermore, numerical simulations for the empirical miss-identification and false identification error rates are provided for finite length codes. This allows us to quantify the scale of error reduction in terms of the codeword length.
% \end{abstract}
% ---
% ---
% \IEEEpeerreviewmaketitle
% ---
\section{Introduction}
\label{Sec.Introduction}
Molecular communication (MC) is a new  paradigm in communication engineering where information is transmitted via signaling molecules \cite{NMWVS12,FYECG16}. Over the past decade, synthetic MC has been extensively studied in the literature from different perspectives including channel modeling \cite{Jamali19}, modulation and detection design \cite{Kuscu19}, biological building blocks for transceiver design \cite{Soldner20}, and information theoretical performance characterization \cite{Gohari16,Rose19}. Moreover, several proof-of-concept implementations of synthetic MC systems have been reported in the literature, see, e.g., \cite{Farsad17,Giannoukos17,Unterweger18}.
%This scheme is widely studies from both theoretical and practical aspects. In particular, in all of existing works on MC, it is assumed that the Shannon communication scheme is the primal and governing task for such MC systems. Information theoretical challenges are discussed in \cite{Gohari16}. A complete survey of MC is conducted in \cite{FYECG16}, a survey for channel modeling is addressed in \cite{Jamali19}, the problem of authentication which is close to identification problem is considered in \cite{Zafar19}, and a full study of fundamentals of MC systems and channel estimation is provided in \cite{Jamali19_2}.

Despite the recent theoretical and technological advancements in the field of  MCs, the fundamental channel capacity limits of most MC systems are still unknown. A mathematical foundation for information theoretical analysis of diffusion-based MC is established in \cite{Hsieh13} where a channel coding theorem is proved.
%, i.e., equivalence of information rate capacity to the code rate capacity, is proved and also it is shown that the diffusion-based molecular channel is stationary, ergodic, and asymptotically decreases the input memory. ==> it is not really clear to the reader!
The information rate capacity of diffusion-based MC was studied in \cite{Pierobon12} where both channel memory and molecular noise are taken into account.
% where the authors utilized thermodynamics and derived the information rate capacity in a closed form style with a receiver of Poisson noise type. ==> This is debatable! let's report something more general about this work
For diffusion-based MC, the capacity limits of molecular timing channels are investigated in \cite{Farsad18} and lower and upper bounds on the corresponding capacity are reported. In \cite{Farsad20}, a new characterization of capacity limits and capacity achieving distributions for the particle-intensity channel are studied. Capacity bounds on point-to-point communication are studied in \cite{Rose19} and a corresponding mathematical framework is established. A comprehensive overview of mathematical challenges and relevant mathematical tools for studying molecular channels is provided in \cite{Gohari16}.

Various applications of MC within the framework of sixth generation wireless networks (6G) \cite{6G+,6G_PST} are associated with event-triggered systems, where Shannon's message transmission capacity, as considered in \cite{Gohari16,Farsad18,Rose19,Pierobon12,Farsad20,Hsieh13}, may not be the appropriate performance metric. In particular, in event-detection scenarios, where the receiver wishes to decides about the occurrence of a specific event in terms of a reliable Yes\,/\,No answer, the so-called identification capacity is the relevant performance measure \cite{AD89}. Specific examples of the identification problem in the context of MC can be found in targeted drug delivery \cite{Muller04,Nakano13} and cancer treatment \cite{Jain99,Wilhelm16,Hobbs_ea98}, where, e.g., a nano-device's objective may be to identify whether or not a specific cancer biomarker exists around the target tissue; in health monitoring \cite{Nakano14,ghavami2020anomaly} where, e.g., one may be interested in whether or not the pH value of the blood exceeds a critical threshold; in natural pheromone communications \cite{wyatt2003pheromones,kaupp2010olfactory} where, e.g., a male insect searches for sex pheromones indicating the presence of a nearby female insect; etc. For these tasks, the identification capacity is deemed to be a key quantitative measure \cite{6G+,6G_PST}. Motivated by this discussion, this paper focuses on the problem of identification in the context of MC systems.

\subsection{Related Work on Identification Capacity}
In Shannon's communication paradigm \cite{S48}, a sender, Alice, encodes her message in a manner that will allow the receiver, Bob, to reliably recover the message. In other words, the receiver's task is to determine which message was sent. In contrast, in the identification setting, the coding scheme is designed to accomplish a different objective \cite{AD89}. The decoder's main task is to determine whether a \emph{particular} message was sent or not, while the transmitter does not know which message the decoder is interested in. Ahlswede and Dueck \cite{AD89} introduced a randomized-encoder identification (RI) scheme, in which the codewords are tailored according to their corresponding random source (distributions). It is well-known that such distributions do not increase the transmission capacity for Shannon's message transmission task \cite{A78}. On the other hand, Ahlswede and Dueck \cite{AD89} established that given local randomness at the encoder, reliable identification is accomplished with a codebook size that is double-exponential in the codeword length $n$, i.e., $\sim 2^{ 2^{nR}}$ \cite{AD89}, where $R$ is the coding rate. This behavior differs radically from the conventional message transmission setting, where the codebook size grows only exponentially, with the codeword length, i.e., $\sim{2^{nR}}$. Therefore, RI yields an exponential gain in the codebook size compared to the transmission problem.

Other non-standard properties of the RI capacity compared to the message transmission capacity can be observed for the discrete memoryless channel (DMC) with feedback, namely, strictly causal feedback from the receiver to the transmitter \emph{can} increase the identification capacity of a DMC \cite{feedback}, but not the message transmission capacity \cite{S56}. Furthermore, for the compound wiretap channel \cite{BD18_2}, the secure RI capacity fulfills a dichotomy theorem that is in strong contrast to the transmission capacity, namely, when the secure RI capacity is greater than zero a price is not paid for secure identification, i.e., the secure RI capacity coincides with the RI capacity. Other unusual behaviors of the RI capacity, in terms of computability and continuity for a correlation-assisted DMC are reported in \cite{Boche19_2}. For instance, the identification capacity for the DMC is not Turing computable. Also, it cannot be represented as the maximization of a continuous function. The construction of RI codes is considered in \cite{VK93,KT99,Bringer09,Bringer10}. Nevertheless, it can be difficult to implement RI codes. Therefore, from a practical point of view, it is of interest to consider the case where the codewords are not selected based on a distribution but rather by means of a deterministic mapping from the message set to the input space. In the literature, this approach is also referred to as identification without randomization \cite{AN99} or deterministic identification (DI) \cite{Salariseddigh_ITW,Salariseddigh_ICC,Salariseddigh_IT}.

In the deterministic coding setup for identification, for DMCs, the codebook size grows only exponentially in the codeword length, similar to the conventional transmission problem \cite{AD89,AN99,Salariseddigh_ICC,J85,Bur00}. However, the achievable identification rates are significantly higher compared to the transmission rates \cite{Salariseddigh_ICC,Salariseddigh_IT}. Deterministic codes often have the advantage of simpler implementation and simulation \cite{Brakerski20,PP09} and explicit construction \cite{A09}. In our recent works \cite{Salariseddigh_ITW,Salariseddigh_ICC,Salariseddigh_IT}, we have considered DI for channels with an average power constraint, including DMCs and Gaussian channels with fast and slow fading, respectively. In the Gaussian case, we have shown that the codebook size scales as $2^{(n\log n)R}$, by deriving bounds on the DI capacity. Furthermore, DI for Gaussian channels is also studied in \cite{Labidi21,Wiese22,BV00,Salariseddigh_ITW}.

\subsection{Contributions}
%Research on micro-scale molecular technology, such as intra-body networks, is still in its early stages and faces many challenges. Nonetheless, MC is a promising contender for future applications, such as the sixth generation of cellular communication (6G) \cite{6G+,6G_PST,Viswanathan20,Tomkos20}, and nanomedical applications including cancer treatment \cite{Hobbs_ea98,Jain99,Wilhelm16} and targeted drug delivery \cite{Muller04,Nakano13}. In MC context, one of the basic models for medium between transmitter and receiver is the Poisson channel. The Poisson channel is also a useful model for an optical communication link with a direct-detection receiver \cite{Fillmore69,Gagliardi76,Shamai90,Shapiro09,Wyner88_I,Wyner88_II,Massey81,Verdu99}. This channel is relevant for practical 6G networks in the context of MC \cite{Gohari16} and optical communications \cite{Cao13,Mceliece81}. The input-output Poisson statistics reads $Y \sim \text{Pois} \left( \lambda + X \right)$ where $X$ is mean of the observed molecules and $\lambda$ stands for expected number of interfering molecules from external and undesired sources.
%,Wyner88_I,Wyner88_II,Massey81,

In this work, we consider MC systems employing molecule counting receivers, where the received signal has been shown to follow the Poisson distribution\footnote{\,The discrete-time Poisson channel (DTPC) is also a useful model for optical communication systems with direct-detection receivers \cite{Gagliardi76,Shamai90,Verdu99}.}\,when the number of released molecules is large, see \cite[Sec.~IV]{Jamali19}, \cite{yilmaz2014arrival,aminian2015capacity} for details. To the best of the authors' knowledge, the identification capacity of the DTPC has not been studied so far. In particular, we consider DI over a DTPC under average and peak power constraints that account for the limited molecule production\,/\,release rate of the transmitter. By deriving positive bounds on the DI capacity of the DTPC, we establish that the codebook size for deterministic encoding scales as $2^{(n\log n)R}$.

The approach to derive the bounds on the capacity is similar to that for Gaussian channels \cite{Salariseddigh_ITW}, namely, to obtain the lower bound we exploit the existence of an appropriate sphere packing within the input space where the mutual distance between the centers of the spheres does not fall below a certain value, and for the upper bound, we assert a certain minimum distance between the codewords of any given sequence of codes with vanishing error probabilities. However, the analysis and the upper bound for the DTPC are different. Here, in the achievability proof, we consider the packing of hyper spheres with radius $\sim n^{\frac{1}{4}}$ inside a larger hyper cube. While the radius of the small spheres in the Gaussian case \cite{Salariseddigh_ITW} tends to zero, here the radius grows in the codeword length, $n$. Yet, we show that we can pack a super-exponential number of spheres within the larger cube. In the converse part of the proof for the DTPC, the derivation of the upper bound on the capacity is more involved compared to that for Gaussian channels \cite{Salariseddigh_ITW} and leads to a larger upper bound. Instead of establishing a minimum distance between the codewords (codeword-wise distance), as in the Gaussian case \cite{Salariseddigh_ITW}, we use a criterion imposed on the symbols of every two codewords, namely, we show that for each pair of different codewords, there exists at least one index for which the ratio of the corresponding symbols is different from 1 (symbol-wise distance). 

The enlarged codebook size of the identification problem compared to the transmission problem may have interesting implications for MC system design. To elaborate on this, we consider two general categories of codes, where the channel uses within each codeword are realized in different manners, namely spatial and temporal codes. For spatial codes, a different type of signaling molecule is used for each channel use. Such a setup can be used to model molecule-mixture communications in mammalian and insect olfactory systems, where a given mixture of different types of molecules represents a {codeword} \cite{Buck05,Kaupp10,Su09}. Thereby, the results presented in this paper may shed light on the extremely large identification capability of natural olfactory systems. In contrast, for temporal codes, different channel uses are realized by releasing the same type of molecules in different time instances. On the one hand, the hardware complexity for temporal codes is lower since only one type of molecule is needed; on the other hand, the receiver has to be equipped with memory to store and jointly process the observations of all channel uses within one codeword. Temporal codes may find applications in synthetic MC identification systems used in, e.g., targeted drug delivery and environmental monitoring \cite{Jamali19,Soldner20}.

Although our theoretical results target the capacity of the DI channel in the standard asymptotic definition, i.e., as $n\to\infty$, we also provide numerical simulations for finite size codes. In particular, our numerical simulations evaluate the empirical miss-identification (type I) and false identification (type II) error rates of the proposed achievable scheme. These results allow us to quantify the error reduction in terms of the codeword length and shed light on the error behavior as a function of the codeword length.

\subsection{Organization}
The remainder of this paper is structured as follows. In Section~\ref{Sec.SysModel}, scenarios for application of DI in the context of MC are discussed and the required preliminaries regarding DI codes are established. Section~\ref{Sec.Results_Analysis} provides the main contributions and results on the message identification capacity of the DTPC. Section~\ref{Sec.Simulation} presents simulation results for the empirical type I and type II error rates. Finally, Section~\ref{Sec.Summary} of the paper concludes with a summary and directions for future research.

\subsection{Notations}

We use the following notations throughout this
paper: Calligraphic letters $\X,\Y,\Z,\ldots$ are used for finite sets. Lower case letters $x,y,z,\ldots$ stand for constants and values of random variables, and upper case letters $X,Y,Z,\ldots$ stand for random variables. Lower case bold symbol $\fx$ indicates a row vector of size $n$, that is, $\fx = (x_1, \dots, x_n)$. Bold symbol $\boldsymbol{1}_n$ indicates the all-one row vector of size $n$. The distribution of a random variable $X$ is specified by a probability mass function (pmf) $p_X(x)$ over a finite set $\X$. All logarithms and information quantities are for base $2$. The set of consecutive natural numbers from $1$ through $M$ is denoted by $[\![M]\!]$. The set of whole numbers is denoted by $\mathbb{N}_{0} \triangleq \{0,1,2,\ldots\}$. The gamma function for non-positive integer $x$ is denoted by $\Gamma(x)$ and is defined as $\Gamma (x) = (x-1) !$, where $(x-1)! \triangleq (x-1) \times (x-2) \times \dots \times 1$. We use the small O notation, $f(n) = o(g(n))$, to indicate that $f(n)$ is dominated by $g(n)$ asymptotically, that is, $\lim_{n\to\infty} \frac{f(n)}{g(n)} = 0$. The big O notation, $f(n) = \mathcal{O}(g(n))$, is used to indicate that $|f(n)|$ is bounded above by $g(n)$ (up to constant factor) asymptotically, that is, $\limsup_{n\to\infty} \frac{|f(n)|}{g(n)} < \infty$. We use the big Omega notation, $f(n) = \Omega(g(n))$, to indicate that $f(n)$ is bounded below by $g(n)$ asymptotically, that is, $g(n) = \mathcal{O}(f(n))$. The $\ell_2$-norm and $\ell_{\infty}$-norm are denoted by $\norm{\mathbf{x}}$ and $\norm{\mathbf{x}}_{\infty}$, respectively. Furthermore, we denote the $n$-dimensional hyper sphere of radius $r$ centered at $\fx_0$ with respect to the $\ell_2$-norm by $\S_{\fx_0}(n,r) = \{\fx\in\mathbb{R}^n : \norm{\fx-\fx_0} \leq r \}$. An $n$-dimensional cube with center $(\frac{A}{2},\ldots,\frac{A}{2})$ and a corner at the origin, i.e., $\mathbf{0} = (0,\ldots,0)$, whose edges have length $A$ is denoted by $\Q_{\f0}(n,A) = \{\fx \in \mathbb{R}^n : 0 \leq x_t \leq A, \forall \, t\in[\![n]\!] \}$.
% In the continuous case, we use the cumulative distribution function $F_X(x)=\Pr(X\leq x)$ for $x\in\mathbb{R}$, or alternatively, the probability density function (pdf) $f_X(x)$, when it exists.

% -------------------------------------------
\section{System Model, MC Scenarios for DI, and Preliminaries}
\label{Sec.SysModel}

In this section, we present the adopted system model, introduce MC scenarios for DI, and establish some preliminaries regarding DI coding.

\subsection{System Model}\label{sec:sysmodel}

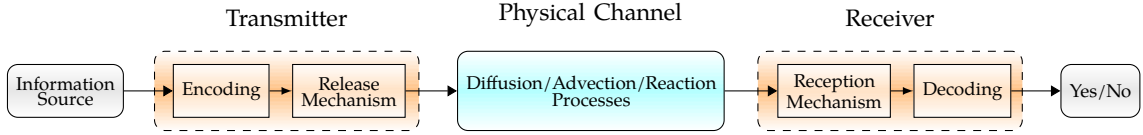
\begin{figure}[!t]
    \centering
    % \captionsetup{justification=raggedright,singlelinecheck=false}
	\scalebox{1}{
\tikzstyle{0} = [draw, -latex']
\tikzstyle{Block_M} = [draw, top color = white, middle color = gray!20, rectangle, rounded corners, minimum height=2em, minimum width=1em]
\tikzstyle{Block1} = [draw, dashed, top color = white, middle color = orange!50, rectangle, rounded corners, minimum height=3em, minimum width=10em]
\tikzstyle{Block2} = [draw, top color = white, middle color = cyan!30, rectangle, rounded corners, minimum height=3em, minimum width=6em]
\tikzstyle{Block3} = [draw, dashed, top color = white, middle color = orange!50, rectangle, rounded corners, minimum height=3em, minimum width=10em]
\tikzstyle{Block1_1} = [draw, top color = white, middle color = orange!20, rectangle, minimum height=2em, minimum width=3em]
\tikzstyle{Block2_1} = [draw, top color = white, middle color = orange!20, rectangle, minimum height=2em, minimum width=3em]
% ---
% \begin{tikzpicture}[remember picture, overlay]
\begin{tikzpicture}
    \node[Block_M] (M) at (-7.35,0) {$\substack{\text{Information} \\ \text{Source}}$};
    \node[Block1, right=.4cm of M] (TX) {};
    \node[above = .2cm of TX] (TX_Text) {\text{\small Transmitter}};
    \node[Block1_1, right=.65cm of M] (enc) {$\substack{\text{Encoding}}$};
    \node[Block1_1, right=.3cm of enc] (particle_Gen) {$\substack{\text{Release} \\ \text{Mechanism}}$};
    \node[Block2, right=.5cm of TX] (channel) {$\substack{\text{Diffusion/Advection/Reaction} \\ \text{Processes}}$};
    \node[above = .2cm of channel] (ch_Text) {\text{\small Physical Channel}};
    \node[Block3, right=.435cm of channel] (RX) {};
    \node[above = .2cm of RX] (RX_Text) {\text{\small Receiver}};
    \node[Block_M, right=.5cm of RX] (YN) {$\substack{\text{Yes/No}}$};
    \node[Block2_1,right= .7cm of channel] (rec) {$\substack{\text{Reception} \\ \text{Mechanism}}$};
    \node[Block2_1,right= .3cm of rec] (ver) {$\substack{\text{Decoding}}$};
    \draw[-Triangle] (M) -- (enc);
    \draw[-latex] (enc) -- (particle_Gen);
    \draw[-Triangle] (particle_Gen) -- (channel);
    \draw[-Triangle] (channel) -- (rec);
    \draw[-latex] (rec) -- (ver);
    \draw[-Triangle] (ver) -- (YN);
\end{tikzpicture}
}
% ---
	\caption{End-to-end chain of communication setup for DI in a generic MC system. Receiver declares a Yes\,/\,No in a reliable manner.}
	\label{Fig.E2E_Chain}
\end{figure}

We focus on an identification setup, where the decoder wishes to reliably determine whether or not a particular message was sent by the transmitter, while the transmitter does not  know which message the decoder is interested in, see Figure~\ref{Fig.E2E_Chain}.
%Example application scenarios for the identification task in the context of MCs are provided in Section~\ref{sec:scenario}. 
To achieve this objective, we establish a coded communication between the transmitter and the receiver over $n$ channel uses of an MC channel.
We consider a stochastic release model, see Figure~\ref{Fig.TX}, where for the $t$-th channel use, the transmitter releases molecules with rate $x_t$ (molecules/second) over a time interval of $T_{\,\text{rls}}$ seconds into the channel \cite{Gohari16}. These molecules propagate through the channel via diffusion and/or advection, and may even be degraded in the channel via enzymatic reactions \cite{Jamali19}. We assume a counting-type receiver which is able to count the number of received molecules. Examples include the transparent (perfect monitoring or passive) receiver, which counts the molecules at a given time within its sensing volume \cite{Unterweger18}, the fully absorbing (perfect sink) receiver, which absorbs and counts the molecules hitting its surface within a given time interval \cite{Yilmaz14_2}, and the reactive (ligand-based) receiver which counts the number of molecules bound to the ligand proteins on its sensing surface at a given time \cite{Ahmadzadeh16}.
% ---
% ---
Assuming that the release, propagation, and reception of individual molecules are statistically similar but independent of each other, the received signal follows Poisson statistics when the number of released molecules is large, i.e., $x_t T_{\,\text{rls}} \gg 1$ \cite[Sec.~IV]{Jamali19}.
Let $X\in\mathbb{R}_{\geq0}$ and $Y\in\mathbb{N}_0$ denote random variables (RVs) modeling the rate of molecule release by the transmitter and the number of molecules observed at the receiver, respectively. The input-output relation for the DTPC is given as follows
% ---
\begin{align}
    \label{Eq.Poisson_Model}
    Y = \text{Pois}\left( \rho X + \lambda \right) \;,\,
\end{align}
% ---
where $\rho X$ is the mean number of observed molecules due to release by the transmitter, $\rho=p_{\,\text{ch}} T_{\,\text{rls}}$, and $p_{\,\text{ch}}\in (0,1]$ denotes the probability that a given molecule released by the transmitter is observed at the receiver. The value of $p_{\,\text{ch}}$ depends on the propagation environment (e.g., diffusion, advection, and reaction processes) and the reception mechanism (e.g., transparent, absorbing, or reactive receiver) as well as the distance between transmitter and receiver, see \cite[Sec.~III]{Jamali19} for the characterization of $p_{\,\text{ch}}$ for various setups and Figure~\ref{Fig.RX} for an example illustration of reception process via absorbing receiver. Moreover, $\lambda\in \mathbb{R}_{> 0}$ is the mean number of observed interfering molecules originating from external noise sources which employ the same type of molecule as the considered MC system.  

The letter-wise conditional distribution of the DTPC output is given by
% ---
\begin{align}
    W(y|x) = \frac{e^{-(\rho x+\lambda)}(\rho x+\lambda)^y}{y!} \,.\,
\end{align}
% ---
Standard transmission schemes employ strings of letters (symbols) of length $n$, referred to as codewords, that is, the encoding schemes use the channel in $n$ consecutive times to transmit one message. As a consequence, the receiver observes a string of length $n$, referred to as output vector (received signal). We assume that different channel uses are orthogonal. This assumption is justified for different MC scenarios in Section~\ref{sec:scenario}. Therefore, for $n$ channel uses, the transition probability law reads
% ---
\begin{align}
    \label{Eq.Poisson_Channel_Law}
    W^n(\fy|\fx) & = \prod_{t=1}^n W(y_t|x_t) = \prod_{t=1}^n \frac{e^{-(\rho x_t+\lambda)}(\rho x_t+\lambda)^{y_t}}{y_t!} \;,\,
\end{align}
% ---
where $\fx = (x_1,\dots,x_n)$ and $\fy = (y_1,\dots,y_n)$ denote the transmit codeword and the received signal, respectively. The codewords are subject to peak and average power constraints as follows
% ---
\begin{align}
    \label{Ineq.Const_X}
    0 \leq x_{t} \leq P_{\,\text{max}} \quad \text{and} \quad \frac{1}{n}\sum_{t=1}^{n} x_{t} \leq P_{\,\text{avg}} \,,\,
\end{align}
% ---
respectively, $\forall t\in[\![n]\!]$, where $P_{\,\text{max}}, P_{\,\text{avg}} > 0$ constrain the rate of molecule release per channel use and over the entire $n$ channel uses in each codeword, respectively. We note that while the average power constraint for the Gaussian channel is a non-linear (square) function of the symbols (signifying the signal energy), here for the DTPC, it is a linear function (signifying the number of released molecules)~\cite{Gohari16}.
% ---% ---
\begin{figure}[!t]
    \centering
	\scalebox{1.1}{
\begin{tikzpicture}
\tikzstyle{Block1_1} = [draw, top color = white, middle color = cyan!35!gray!20, rectangle, minimum height=2em, minimum width=3.3em]
\tikzstyle{Block1_2} = [draw, top color = yellow!45!gray!10, bottom color = yellow!45!gray!10, rectangle, minimum height=21mm, minimum width=32mm]
\foreach \a in {1.2}{
\node[Block1_1] (enc) at (\a,0) {$\text{\small Encoder}$};
\node[Block1_2, right = 1 cm of enc] (particle_Gen) {};
\node[above = .2cm of particle_Gen] (PS_Text) {\text{\small Particle Storage}};
\draw[-latex] (enc) -- node[above]{$x_t$} (particle_Gen);
\draw[|-|, very thick, mycolor3] (particle_Gen.-10) -- (particle_Gen.10);
\draw[dashed,-latex] (\a+1.2,0) -- (\a+1.2,-1.5) -- (\a + 5.3,-1.5) -- (\a + 5.3,0) -- (particle_Gen.0);}
\draw plot [only marks, mark options={color=black}, mark=*, mark size = .7pt , domain = 3:6, samples=1000] (\x,{2*rnd-1});
\end{tikzpicture}
}
	\caption{Transmission configuration in a Poisson concentration type. The value of the channel input, $x_t$  controls the outlet size bio-chemically where it adjusts the width of the gate, that is, $x_t=0$ represents a closed gate and $x_t=A$ accounts to a maximal opening. At every time instance $t$, the width of the open gate, marked in blue, increases with the input value $x_t$. The exact number of emitted particles is modelled by a Poisson distribution with mean of $x_t T_{\,\text{rls}}$ for $T_{\,\text{rls}}$ to be the symbol interval.}
	\label{Fig.TX}
\end{figure}
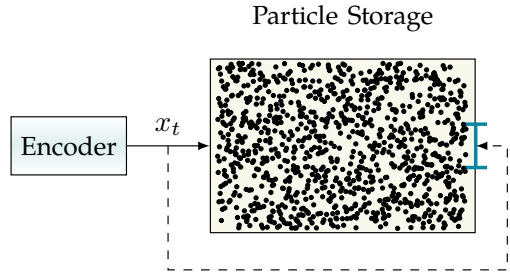
% ---
\begin{figure}[!b]
    \centering
	\tikzset{%
  >=latex, % option for nice arrows
  inner sep=0pt,%
  outer sep=2pt,%
  mark coordinate/.style={inner sep=0pt,outer sep=0pt,minimum size=3pt,
    fill=black,circle}%
}
\usetikzlibrary{shapes.misc}

\tikzset{cross/.style={cross out, draw=black, minimum size=2*(#1-\pgflinewidth), inner sep=0pt, outer sep=0pt}, cross/.default={1pt}}

\begin{tikzpicture} % "THE GLOBE" showcase
\pgfmathsetseed{116}
\def\R{2} % sphere radius
\def\angEl{35} % elevation angle
\filldraw[ball color = yellow!2!cyan!5,opacity = .09] (0,0) circle (\R);
\foreach \t in {-80,-60,...,80} { \DrawLatitudeCircle[\R]{\t} }
\foreach \t in {-5,-35,...,-175} { \DrawLongitudeCircle[\R]{\t} }
\node [] at (0,2.3) {\text{\small Spherical Receiver}};
% --- Random Walk
\draw [blue!60!gray,fill] (-8,0) circle [radius = 0.05];
\draw [blue!60!gray,fill] (1,.5) circle [radius = 0.05] node [right] {};
\node [] at (-8,.3) {\text{\small Information Molecule}};
% \draw [red, very thick] (1,.5) node [cross, rotate=10, ,inner sep=0pt, minimum size=8pt] {};
% \draw [->]
%   {decorate[decoration={random steps, segment length=8,amplitude=10}]
%   {decorate[decoration={random steps, segment length=8,amplitude=10}]
%   {decorate[decoration={random steps, segment length=8,amplitude=10}]     
%   { (-5,0) to[out=45,in=0] ++ (0,-1) -- (1,-3) -- (3,-1) to[out=90,in=-90] (1,.5) }}}};
\draw [-latex,thin]
  {decorate[decoration={random steps, segment length=4,amplitude=5}]
  {decorate[decoration={random steps, segment length=4,amplitude=5}]
  {decorate[decoration={random steps, segment length=4,amplitude=5}]     
  { (-8,0) to[out=0,in=0] ++ (0,-1) -- (1,-2.5) -- (2.5,-1) to[out=0,in=0] (1,.5) }}}};
% ---
\end{tikzpicture}
	\caption{Trajectory of captured particles within the surface of an absorbing receiver. Each small propagation step is governed by Brownian motion. Once the information particles hits the absorbing boundary, they will turn in-reversibly into another type of molecule. As a result, such hitting particles will immediately disappear from the environment.}
	\label{Fig.RX}
\end{figure}
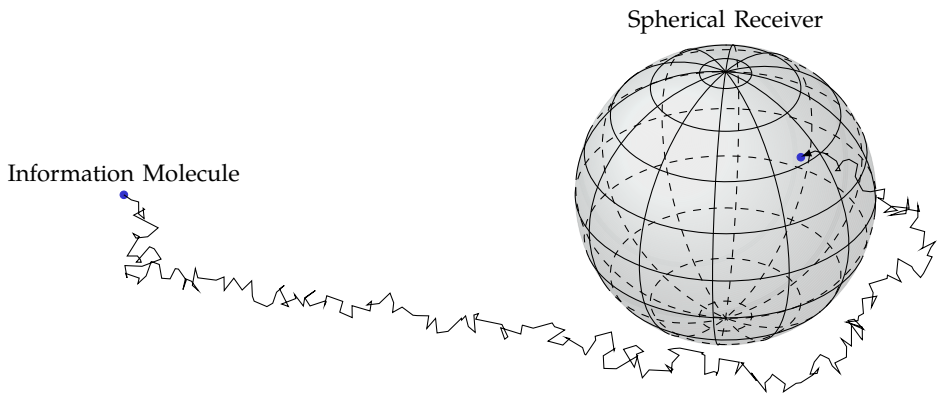
% ---
\subsection{Spatial vs. Temporal Channel Uses}
\label{sec:scenario}
The coding scheme proposed in this paper requires $n$ independent uses of the MC channel; however, how the MC channel is accessed for each channel use may depend on the application of interest. In the following, we introduce two application scenarios, which employ spatial and temporal channel uses, respectively.

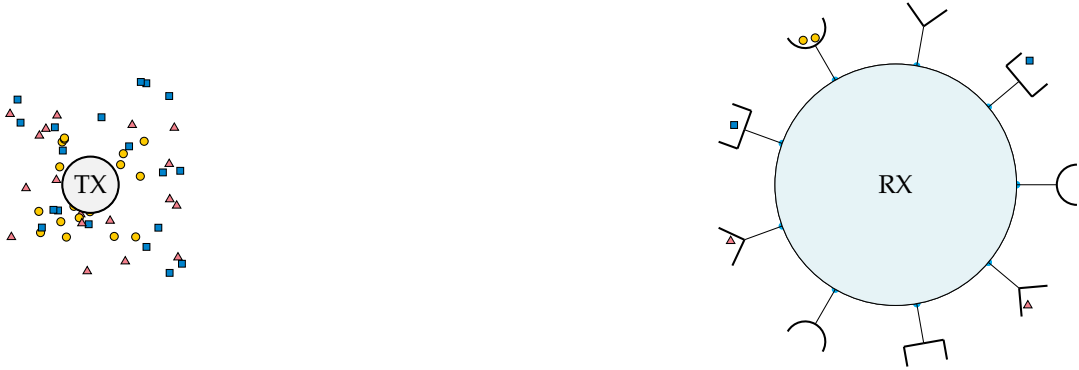
\begin{figure}[tb]
	\centering
	\resizebox{.9\linewidth}{!}{\tikzset{semicircle/.pic = {
    \draw (0:3) -- (0:4);
    \draw [thick, xshift = .5cm, yshift = .5cm](0*40:4) arc (90:270:.5);
}}
% ---
\tikzset{triangle/.pic = {
    \draw (0:3) -- (0:4);
    \draw [thick] (0:4) -- ([xshift = .5cm, yshift = .5cm]0:4);
    \draw [thick] (0:4) -- ([xshift = .5cm, yshift = -.5cm]0:4);
}}
% ---
\tikzset{square/.pic = {
    \draw (0*40:3) -- (0*40:4);
    \draw [thick] (0:4) -- ([xshift = 0cm, yshift = .5cm]0:4);
    \draw [thick] (0:4) -- ([xshift = 0cm, yshift = -.5cm]0:4);
    \draw [thick] ([xshift = 0cm, yshift = +.5cm]0:4) -- ([xshift = .5cm, yshift = +.5cm]0:4);
    \draw [thick] ([xshift = 0cm, yshift = -.5cm]0:4) -- ([xshift = .5cm, yshift = -.5cm]0:4);
}}

\tikzset{triangle/.style = { isosceles triangle,
    isosceles triangle apex angle=60,
    draw,fill= rot_9b!50, minimum size = .1cm, rotate=90,
    inner sep=0pt
    }}
% ---
% \tikzset{RX/.pic = {
    % \foreach \x [count=\p] in {0,...,11} {
    % \node [shape = circle, fill = black, scale = 0.3] (\p) at (-\x*40:2) {};},
    % \foreach \x [count=\p] in {0,3,6} {\path[rotate = -\x*40, transform shape] pic{semicircle};}
    % \foreach \x [count=\p] in {1,4,7} {\path[rotate = -\x*40, transform shape] pic{triangle};}
    % \foreach \x [count=\p] in {2,5,8} {\path[rotate = -\x*40, transform shape] pic{square};}
    % \draw (1) arc (0:360:2);
% }}
% ---
\begin{tikzpicture}[scale=0.55]
    % Around TX
    % \pgfmathsetseed{1125}
    \pgfmathsetseed{53}
    \foreach \p in {1,...,20}{
    % \draw [fill = orange_6b] (3*rand-19 , 3*rand+.5) circle (0.1);
    % \draw [fill = blau_2b] (3*rand-19.5 , 3*rand) rectangle ++ (.17,.17);
    % \node[triangle] at (3*rand-18.5 , 3*rand+.5) {};
    \draw [fill = orange_6b] (1.4*rand-20 , 1.4*rand) circle (0.1);
    \node[triangle] at (2.2*rand-20 , 2.2*rand) {};
    \draw [fill = blau_2b] (2.5*rand-19.5 , 2.5*rand) rectangle ++ (.17,.17);
    }
    % Around RX
    \node[triangle] at (-4.1,-1.4) {};
    \node[triangle] at (3.28,-3) {};
    
    \draw [fill = blau_2b] (3.24,3) rectangle ++ (.17,.17);
    \draw [fill = blau_2b] (-4.1,1.4) rectangle ++ (.17,.17);

    \draw [fill = orange_6b] (-2,3.65) circle (0.1);
    \draw [fill = orange_6b] (-2.3,3.59) circle (0.1);
    
    % \node[isosceles triangle,
    % isosceles triangle apex angle=60,
    % draw,fill=teal!10,
    % minimum size =1cm, rotate=90] (T60) at (-20,0){};
    
    \draw [fill = gray!10, fill opacity = 1] (-20,0) circle (0.7 cm);
    \draw [thick] (-20,0) circle (.7cm);
    \node [] at (-20,0) {\text{TX}};
    % RX
    \foreach \x [count=\p] in {0,...,11} {
    \node [shape = circle, fill = cyan, scale = 0.3] (\p) at (-\x*40:3) {};}
    \foreach \x [count=\p] in {0,3,6} {\path[rotate = -\x*40, transform shape] pic{semicircle};}
    \foreach \x [count=\p] in {1,4,7} {\path[rotate = -\x*40, transform shape] pic{triangle};}
    \foreach \x [count=\p] in {2,5,8} {\path[rotate = -\x*40, transform shape] pic{square};}
    \draw (1) arc (0:360:3);
    \draw [fill = mycolor12!10, fill opacity = 1] (0,0) circle (3 cm);
    \node [] at (0,0) {\text{RX}};
\end{tikzpicture}}
	\caption{Illustration of an olfactory-inspired MC system, where three orthogonal molecule types, namely type square, triangle, and circle, are shown. The transmitter secretes a mixture of these molecules corresponding to a particular message. Each receptor located on the receiver surface is sensitive to only one type of molecule. The receiver's task is to determine whether or not a desired message (molecular mixture) has been sent by the transmitter.}
	\label{Fig.Molecular_Othogonality}
\end{figure}

\textbf{Spatial channel use:} The mammalian olfactory system is believed to possess the capacity of discriminating many thousands of different chemical mixtures, where each chemical mixture is often associated with a specific event or an external object \cite{ Buck92, Buck05,Bushdid14}. Examples of such mixtures include mating pheromones, social pheromones, food odorants, repellent odorants (e.g., implying toxic materials or predators), etc. Hence, the \emph{Ahlswede-Dueck} identification problem applies to natural olfaction since each molecular mixture conveys a particular message that the receiver may be interested in. Here, each molecular mixture can be seen as a codeword, where the channel is accessed spatially and simultaneously via the different types of molecules in the mixture. The extraordinary discriminatory power of natural olfactory systems has motivated researchers to develop biosensors that mimic the structural and functional features of natural olfaction \cite{pearce2006handbook,barbosa2018protein}. Motivated by this, we consider a communication scenario, where the transmitter releases a mixture of $n$ different types of molecules to convey a message to the receiver, see Figure~\ref{Fig.Molecular_Othogonality}. The receiver is equipped with a dedicated type of receptor for each type of molecule, which ensures the orthogonality of the $n$ channel uses. The receiver's task is to determine whether or not a desired message (molecular mixture) has been sent by the transmitter.
% ---
% ---

\textbf{Temporal channel use:} For spatial channel uses, the complexity of transmitter and receiver may be high as they have to be able to generate and detect $n$ different types of molecules, respectively. To avoid this complexity, one may employ only one type of molecule and access the MC channel at different time instances. Thereby, transmitter and receiver have to be equipped with memory for generation and processing of all $n$ channel uses, respectively. In addition, due to the dispersive nature of the diffusive MC channel, the channel has memory and proper measures have to be taken to ensure the orthogonality of different channel uses. An immediate approach is to make the symbol duration sufficiently large such that the channel response (practically) decays to zero within each symbol interval. However, this may lead to an inefficient design due to the reduction of the rate of channel access. More efficient approaches proposed in the literature include the use of enzymes \cite{Noel14} and reactive cleaning molecules \cite{Jamali18} to generate a concentrated channel response for the desired signaling molecules, see, e.g., \cite[Fig.~15]{Jamali19}.

\subsection{DI Coding for the DTPC}

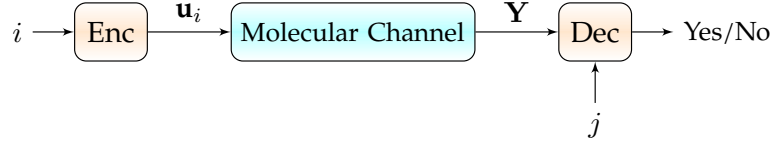
\begin{figure}[!t]
    \centering
	\scalebox{1.1}{
\tikzstyle{l} = [draw, -latex']
\tikzstyle{Block1} = [draw,top color=white, middle color=white!80!orange, rectangle, rounded corners, minimum height=2em, minimum width=2.5em]
\tikzstyle{Block2} = [draw,top color=white, middle color=cyan!30, rectangle, rounded corners, minimum height=2em, minimum width=8em]
\tikzstyle{Block3} = [draw,top color=white, middle color=cyan!10, rectangle, rounded corners, minimum height=2em, minimum width=2em]
\tikzstyle{Block4} = [draw,top color=white, middle color=red!20, rectangle, rounded corners, minimum height=2em, minimum width=2em]
\tikzstyle{Block5} = [draw,top color=white, middle color=white!80!orange, rectangle, rounded corners, minimum height=2em, minimum width=2.5em]
\tikzstyle{input} = [coordinate]
\tikzstyle{sum} = [draw, circle,inner sep=0pt, minimum size=5mm,  thick]
\tikzstyle{arrow}=[draw,->]
\tikzstyle{small_node} = [draw, circle,inner sep=0pt, minimum size=.8mm,thick]
\begin{tikzpicture}[auto, node distance=2cm,>=latex']
\node[] (M) {$i$};
\node[Block1,right=.5cm of M] (enc) {Enc};
% \node[Block2, right=.7cm of enc] (Pois_Gen) {$\text{\small Diffusion CHL}$};
\node[Block2, right=1cm of enc] (channel) {$\text{\small Molecular Channel}$};
%\node[sum, right=.7cm of channel] (adder) {$+$};
%\node[Block4, right=.5cm of adder] (Pois) {\text{Pois(.)}};
\node[Block5, right=1cm of channel] (dec) {Dec};
\node[below=.5cm of dec] (Target) {$j$};
\node[right=.5cm of dec] (Output) {\text{\small Yes/No}};
%\node[above=.5cm of adder] (noise) {$\lambda$};
%
\draw[->] (M) -- (enc);
\draw[->] (enc) -- node[above]{$\textbf{u}_i$} (channel);
%\draw[->] (Pois_Gen) -- (channel);
%\draw[->] (channel) -- node [above] {} (adder);
%\draw[->] (noise) -- (adder);
%\draw[->] (Pois_Gen) -- node [above] {\text{}} (channel);
%\draw[->] (adder) --node[above]{} (Pois);
\draw[->] (channel) --node[above]{$\fY$} (dec);
\draw[->] (dec) -- (Output);
\draw[->] (Target) -- (dec);
\end{tikzpicture}
}
	\caption{DI scheme for MC with under the DTPC model. Relevant processes in the molecular channel are diffusion, advection, and reaction. $Y_t$ is Poisson distributed with rate $ \rho u_{i,t}+\lambda$. Given the observation $\fY$, decoder asks whether $j$ equals $i$ or not.}
	\label{Fig.PoissonChannel}
\end{figure}

%\textcolor{blue}{For notational simplicity, we define random variable $U=p_{\,\text{ch}} X \in \mathbb{R}_{>0}$ (i.e., expected number of molecules observed at the receiver) and design the code in terms of $U$ instead of $X$. Thereby, we assume that $U$ is a continuous variable which is a reasonable approximation in the context of MCs where the integer $X\in\mathbb{N}_0$ is very large and the channel probability $p_{\,\text{ch}}\in[0,1]$ is quite small.}
The definition of a DI code for the DTPC is given below.
%%%
\begin{definition}[Poisson DI Code]
\label{GdeterministicIDCode}
An $\left( L(n,R),n,\lambda_1,\lambda_2 \right)$ DI code for a DTPC $\W$ under average and peak power constraints of $P_{\,\text{ave}}$ and $P_{\,\text{max}}$, respectively, and for integer $L(n,R)$, where $n$ and $R$ are the codeword length and coding rate, respectively, is defined as a system $(\U,\mathscr{D})$ which consists of a codebook $\U=\{ \mathbf{u}_i \}_{i\in[\![L]\!]} \subset \R^n$, such that
% ---
\begin{align}
    \label{Ineq.Power_Const}
    0 \leq u_{i,t} \leq P_{\,\text{max}} \quad \text{and} \quad \frac{1}{n}\sum_{t=1}^{n} u_{i,t} \leq P_{\,\text{avg}} \;,\,
\end{align}
% ---
$\forall i\in[\![L]\!]$, $\forall t\in[\![n]\!]$, and a collection of decoding regions $\mathscr{D}=\{ \D_i \}_{i\in[\![L]\!]}$ with
% ---
\begin{align*}
    \bigcup_{i=1}^{L(n,R)}\D_i\subset\mathbb{N}_0^n \;.\,
\end{align*}
% ---
Given a message $i\in [\![L]\!]$, the encoder transmits $\mathbf{u}_i$, see Figure~\ref{Fig.PoissonChannel}, the decoder's aim is to answer the following question: Was a desired message $j$ sent or not? There are two types of errors that may occur:
Rejection of the true message or acceptance of a false message. These errors are referred to as type I and type II errors, respectively.\\
The corresponding error probabilities of the identification code $(\U,\mathscr{D})$ are given by
% ---
\begin{align}
    P_{e,1}(i) & = 1 -\sum_{\fy \in \D_i} W^n \left( \fy \, \big| \, \fu_i \right) &&\hspace{-2cm}  \text{(miss-identification error)} \,,\, \label{Eq.GTypeIErrorDef}
    \\
    P_{e,2}(i,j) & = \sum_{\fy \in \D_j} W^n \left( \fy \, \big| \, \fu_i \right) && \hspace{-2cm} \text{(false identification error)} \,.\;
    \label{Eq.GTypeIIErrorDef}
\end{align}
% ---
and satisfy the following bounds
% ---
\begin{align}\label{Eq.ErrorConstraints}
P_{e,1}(i) \leq \lambda_1 \quad \text{and} \quad P_{e,2}(i,j) \leq \lambda_2 \;,\,
\end{align}
% ---
$\forall \, i,j \underset{i\neq j}{\in} [\![L]\!]$ and every $\lambda_1,\lambda_2>0$. A rate $R>0$ is called achievable if for every $\lambda_1,\lambda_2>0$ and sufficiently large $n$, there exists an $(L(n,R),n,\lambda_1,\lambda_2)$ DI code. The operational DI capacity of the DTPC is defined as the supremum of all achievable rates, and is denoted by $\mathbb{C}_{DI}(\W,L)$.
\qed
\end{definition}
% ---
% ---
% ----------------------------
\section{DI Capacity of the DTPC}
\label{Sec.Results_Analysis}

In this section, we first present our main results, i.e., lower and upper bounds on the achievable identification rates for the DTPC. Subsequently, we provide the detailed proofs of these bounds.

\subsection{Main Results}

The DI capacity theorem for the DTPC is stated below.

% ---
\begin{theorem}
\label{Th.PDICapacity}
The DI capacity of the DTPC $\W$ subject to average and peak power constraints of the form $n^{-1} \sum_{t=1}^n u_{i,t} \allowbreak \leq P_{\,\text{ave}}$ and $0 \leq u_{i,t} \leq P_{\,\text{max}}$, respectively, in the super-exponential scale, i.e., $L(n,R)=2^{(n\log n)R}$, is bounded by
% ---
\begin{align}
    \label{Ineq.LU}
    \frac{1}{4} \leq \mathbb{C}_{DI}(\W,L) \leq \frac{3}{2} \,.\;
\end{align}
% ---
% Hence, the DI capacity is infinite in the exponential scale  and zero in the double exponential, i.e.,
% %%%
% \begin{align}
%     \label{Eq.PDICapacity2}
%     \mathbb{C}_{DI}(\W,L) = 
%     \begin{cases}
%     \infty & \text{ for 
% 	$L(n,R)=2^{(n\log n)R}$.}\\
% 	0&\text{ for 
% 	$L(n,R)=2^{2^{(n\log n)R}}$.}
% \end{cases}
% \end{align}
% %%%
\end{theorem}
%%%
% ---
\begin{IEEEproof}
The proof of Theorem~\ref{Th.PDICapacity} consists of two parts, namely the achievability and the converse proofs, which are provided in Sections~\ref{sec:achievable} and \ref{Subsec.ConvFast}, respectively. 
%The second part of the theorem is a direct consequence of the arguments in \cite[Rem.~3]{Salariseddigh_ITW}.
\end{IEEEproof}
% ---
% \textcolor{red}{Mohammad, can we highlight the  insights obtained from this theorem, before giving the proof? For example, 
% \begin{itemize}
%     \item Describing the scale of capacity and giving (9).   
%     \item Why are the bounds in (7) independent of the power budgets? Interesting insights regarding the asymptotic $P_{\,\text{ave}}, P_{\,\text{max}}\to 0,\infty$.
%     \item The discussion on the adopted decoder and why it is different from that of the Gaussian channel.
% \end{itemize}}
% ---
Before we provide the proof, we highlight some insights obtained from Theorem~\ref{Th.PDICapacity} and its proof.

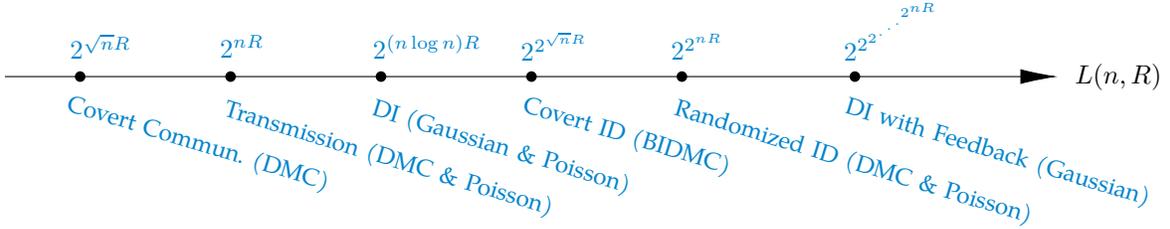
\begin{figure}[!t]
    \centering
    % \vspace{10mm}
    \scalebox{1}{
\begin{tikzpicture}
% [remember picture, overlay]
% \foreach \x in {8}
\tikzmath{\x = 7.5;}

\node (Scale) at (14.8-\x,0) {\text{$L(n,R)$}};
\draw[-{Latex[length=5mm, width=2mm]}] (0-\x,0) -- (14-\x,0);
% ---
\node (CC) at (1-\x,0) {};
\fill [black] (CC) circle (2pt);

\node[rotate = 0] at (1.27-\x,.4){\text{$\textcolor{blau_2b}{2^{\sqrt{n}R}}$}};

\node[rotate = -17] at (2.57-\x,-.9) {\text{\small \textcolor{blau_2b}{Covert Commun. (DMC)}}};
% ---
\node (TR) at (3-\x,0) {};
\fill [black] (TR) circle (2pt);

\node[rotate = 0] at (3.15-\x,.4){\text{$\textcolor{blau_2b}{2^{nR}}$}};

\node[rotate = -17] at (5.1-\x,-1.1) {\text{\small \textcolor{blau_2b}{Transmission (DMC \& Poisson)}}};

% ---
\node (DI) at (5-\x,0) {};
\fill [black] (DI) circle (2pt);

\node[rotate = 0] at (5.61-\x,.4){$\textcolor{blau_2b}{2^{(n\log n)R}}$};

\node[rotate = -17] at (6.6-\x,-.95) {\text{\small \textcolor{blau_2b}{DI (Gaussian \& Poisson)}}};
% ---
\node (CI) at (7-\x,0) {};
\fill [black] (CI) circle (2pt);

\node[rotate = 0] at (7.3-\x,.4){\text{$\textcolor{blau_2b}{2^{2^{\sqrt{n}R}}}$}};

\node[rotate = -17] at (8.27-\x,-.8) {\text{\small \textcolor{blau_2b}{Covert ID (BIDMC)}}};
% ---
\node (RI) at (9-\x,0) {};
\fill [black] (RI) circle (2pt);

\node[rotate = 0] at (9.2-\x,.4){\text{$\textcolor{blau_2b}{2^{2^{nR}}}$}};

\node[rotate = -17] at (11.28-\x,-1.15) {\text{\small \textcolor{blau_2b}{Randomized ID (DMC \& Poisson)}}};

% ---
\node (RI) at (11.3-\x,0) {};
\fill [black] (RI) circle (2pt);

\node[rotate = 0] at (11.8-\x,.6){\text{$\textcolor{blau_2b}{2^{2^{2^{\iddots^{2^{nR}}}}}}$}};

\node[rotate = -17] at (13.2-\x,-1.02) {\text{\small \textcolor{blau_2b}{DI with Feedback (Gaussian)}}};

\end{tikzpicture}
}
    % \vspace{18mm}
    \caption{Spectrum of codebook sizes for different transmission and identification setups. Apart from the conventional exponential and double exponential codebook sizes for transmission \cite{S48} and RI \cite{AD89}, respectively, different non-standard codebook sizes are observed for other communication tasks, such as \emph{covert communication} \cite{Bloch16} or \emph{covert identification} \cite{ZT20} for the binary-input DMC (BIDMC), where the codebook size scales as $2^{\sqrt{n}R}$ and $2^{2^{\sqrt{n}R}}$, respectively. For the Gaussian DI channel with feedback \cite{Labidi21}, the codebook size can be arbitrarily large. In \cite{Wiese22} the result is generalized for channels with non-discrete additive white noise and positive message transmission feedback capacity.}
    \label{Fig.Scales_Spectrum}
\end{figure}

\textbf{Scale}: Theorem~\ref{Th.PDICapacity} shows a different behavior compared to the traditional scaling of the codebook size with respect to codeword length $n$. The bounds given in Theorem~\ref{Th.PDICapacity} are valid in the super-exponential scale of $L=2^{(n\log n)R}$ which is in between the conventional exponential and double exponential codebook sizes (see Figure~\ref{Fig.Scales_Spectrum}). Given such a codebook size it follows \cite[see Rem.~1]{Salariseddigh_ICC} that the capacity values in the standard codebook sizes, i.e., exponential and double exponential, are infinite and zero, respectively.
% i.e., $\mathbb{C}_{DI}(\W,L=2^{nR}) = \infty$ and $\mathbb{C}_{DI}(\W,L=2^{2^{nR}}) = 0$, respectively.

% {\RED $@$ Mohammad: By DMC \& Poisson, do you mean DMPC? Also, if some of the abbreviations are not defined in the text before, then define them in the figure caption!}
% {\BLUE I meant that both channels scale double exponential. For DMC, Ahlswede and Dueck prove it \cite{AD89} and for Poisson (continuous time), Burnashev prove it \cite{BV00}. 
% \RED OK, I understand now, but it is still strange since DMC is not disjoint from Poisson or Gaussian channel.}
% -----------------------------------------
\textbf{Budget for Molecule Release}: The proposed capacity bounds in the super-exponential scale are independent of the values of $P_{\,\text{ave}}$ and $P_{\,\text{max}}$ as long a the codeword length $n$ grows sufficiently large, i.e., $n\to\infty$. However, for finite $n$, the codebook size is indeed a function of $P_{\,\text{ave}}$ and $P_{\,\text{max}}$. This can be readily seen from the achievability proof, where the codebook size in its raw form (see \eqref{Eq.Log_L}) before division by the dominant term reads
% ---
\begin{align}
    L(n,R) = 2^{(n\log n)R  + n ( \log \frac{A}{e\sqrt{a}}) + o(n)} \;.\,
\end{align} 
% ---
where $A=\min \left(P_{\,\text{ave}},P_{\,\text{max}} \right)$ and $a>0$ is a parameter of the codebook construction, cf. \eqref{Eq:epsilon}. In other words, the codebook size increases as $A$ increases; however, since $A$ appears in a term that is exponential in $n$, i.e., $\sim 2^{n ( \log \frac{A}{e\sqrt{a}})}$, the influence of $A$ becomes negligible compared to the dominant super-exponential term, i.e., $2^{(n\log n) R}$ as $n\to\infty$\footnote{\,It is interesting to recall that the codebook size for the transmission capacities of both the DTPC \cite[see Eq.~(5)]{Lapidoth08_2} and the Gaussian channel \cite{Urbanke98,Shannon49} scale with $2^{n\log \sqrt{A}}$ in terms of $A$.}. While the proof of Theorem~\ref{Th.PDICapacity} mainly concerns the asymptotic regime of $n\to\infty$, we are still able to get some insight for finite $n$, too. For instance,  the error constraints in \eqref{Eq.ErrorConstraints} can be met by the proposed achievable scheme even for finite $n$ if $A$ is sufficiently large and $a=\Omega(A^2)$, cf. \eqref{Ineq.TypeI}, \eqref{Ineq.E_0}, and \eqref{Ineq.E_1}. A comprehensive study of the achievable DI rates for finite $n$ constitutes an interesting research topic for future work, but is beyond the scope of this paper.

%To observe behavior of the capacity in the asymptotic region of power constraints, we recall the case 1 and 3 of Equation~(\ref{Eq.n_Cube_Vol}) describes the vanishing regime, i.e., $P_{\,\text{ave}}, P_{\,\text{max}} \to 0$ and large scale regime, i.e., $P_{\,\text{ave}},P_{\,\text{max}} \to \infty$, respectively. For both small or large values of constraints our capacity bounds holds true. For detailed derivation see Appendix.~\ref{App.SP_Diverging_Radius}.

% ---
\textbf{Adopted Decoder}:
For the achievability proof, we adopt a decoder that upon observing an output sequence $\fy$, declares that the message $j$ was sent if the following condition is met
% ---
\begin{align}
    %   \left| \frac{1}{n} \sum_{t=1}^n \left[ \left( y_t - \left( \rho u_{j,t} + \lambda \right) \right)^2 - \left( \lambda + \rho u_{j,t} \right) \right] \right| \leq \delta_n \,.\,
    \left| \, \norm{ \fy - \mathbb{E} \left( \fy \, \big| \, \fu_j \right) }^2 - \mathbb{E} \left( \norm{ \fy - \mathbb{E} \left( \fy | \fu_j \right) }^2 \, \Big| \, \fu_j \right)
    \right| \leq n\delta_n \,.\,
    \label{Eq.Decoding_Set_0}
\end{align}
% ---
where $\fu_j = [u_{j,1},\ldots,u_{j,n}]$ is the codeword associated with message $j$ and $\delta_n$ is a decoding threshold. In contrast to the popular distance decoder used for Gaussian channels \cite{Salariseddigh_ITW} that includes only the distance term $\| \fy - \mathbb{E} ( \fy \,|\, \fu_j ) \|$, the proposed decoder in \eqref{Eq.Decoding_Set_0} comprises the additional correction term $\mathbb{E} \big( \| \fy - \mathbb{E} ( \fy \,|\, \fu_j ) \|^2 \,|\, \fu_j \big)$. This choice stems from the fact that the noise in the DTPC is signal dependent \cite{Jamali19}. Therefore, the variance of $\| \fy - \mathbb{E} ( \fy \,|\, \fu_j ) \|$ depends on the adopted codeword $\fu_j$, which implies that unlike for the Gaussian channel, here the radius of the decoding region is not constant for all codewords.
% ---
% ------------------------
\subsection{Achievability}
\label{sec:achievable}
Consider DTPC $\W$. We show achievability of \eqref{Ineq.LU} using a packing of hyper spheres and a distance decoder. We pack hyper spheres with radius $\sim n^{\frac{1}{4}}$ inside a larger hyper cube. While the radius of the spheres in a similar proof for Gaussian channels vanishes, as $n$ increases \cite{Salariseddigh_ITW}, the radius here diverges to infinity. Yet, we can obtain a positive rate while packing a super-exponential number of spheres satisfying the power and error constraints in \eqref{Eq.GTypeIErrorDef}-\eqref{Eq.ErrorConstraints}. A DI code for the DTPC $\W$ is constructed as follows.
% ------------------------------------
\subsubsection*{Codebook construction}
\label{Subsec.CodebookConstruction_1}
Let
% ---
\begin{align}
    \label{Eq.A}
    A = \min \left(P_{\,\text{ave}},P_{\,\text{max}} \right) \;.\,
\end{align}
% ---
In the following, we restrict ourselves to codewords that meet the condition $0 \leq x_t \leq A$, $\forall \, t \in [\![n]\!]$. We argue that this condition ensures both the average and the peak power constraints in \eqref{Ineq.Const_X}. In particular, when $P_{\,\text{ave}} \geq P_{\,\text{max}}$, then $A = P_{\,\text{max}}$ and the constraint $0 \leq x_t \leq A$ automatically implies that the constraint $\frac{1}{n} \sum x_t \leq P_{\,\text{ave}}$ is met, hence, in this case, the setup with average and peak power constraints simplifies to the case with only a peak power constraint. On the other hand, when $P_{\,\text{ave}} < P_{\,\text{max}}$, then $A = P_{\,\text{ave}}$ and by $0 \leq x_t \leq A$, $\forall \, t \in [\![n]\!]$, both power constraints are met, namely $\frac{1}{n} \sum x_t \leq P_{\,\text{ave}}$ and $0 \leq x_t \leq P_{\,\text{max}}$, $\forall \, t \in [\![n]\!]$. Hence, in the following, we restrict our considerations to a hyper cube with edge length $A$.

We use a packing arrangement of non-overlapping hyper spheres of radius $r_0 = \sqrt{n\epsilon_n}$ in a hyper cube with edge length $A$, where
% ---
\begin{align}
    \label{Eq:epsilon}
    \epsilon_n = \frac{a}{n^{\frac{1}{2}(1-b)}} \;,\,
\end{align}
% ---
and $a>0$ is a non-vanishing fixed constant and $0 < b < 1$ is an arbitrarily small constant.

\begin{figure}[t]
    \centering
	\scalebox{0.92}{
\begin{tikzpicture}[scale=.7][thick]
%\draw[thick] (0,0) circle (3.1cm);
% \tikzstyle{Block2} = [draw,top color=white, bottom color=cyan!30, rectangle, rounded corners, minimum height=2em, minimum width=2em]

%% Entire Spheres
\draw (-1.41,-1.41) circle (1cm);
\draw [fill=mycolor7, fill opacity=0.2] (-1.41,-1.41) circle (1cm);
\draw (0,0) circle (1cm);
\draw [fill=mycolor7, fill opacity=0.2] (0,0) circle (1cm);
\draw (1.41,1.41) circle (1cm);
\draw [fill=mycolor7, fill opacity=0.2] (1.41,1.41) circle (1cm);
% ---
\draw (+.52,-1.93) circle (1cm);
\draw [fill=mycolor7, fill opacity=0.2] (+.52,-1.93) circle (1cm);
\draw (1.93,-.52) circle (1cm);
\draw [fill=mycolor7, fill opacity=0.2] (1.93,-.52) circle (1cm);
% ---
\draw (-1.93,+.52) circle (1cm);
\draw [fill=mycolor7, fill opacity=0.2] (-1.93,.52) circle (1cm);
\draw (-.52,1.93) circle (1cm);
\draw [fill=mycolor7, fill opacity=0.2] (-.52,1.93) circle (1cm);
% --- Partial Spheres
\node [fill=black, shape=circle, inner sep=.4pt] ($.$) at (2.45,-2.45) {};
\draw (2.45,-2.45) circle (1cm);
% \draw [fill=cyan!20!white, fill opacity=0.2] (2.45,-2.45) circle (1cm);
\draw [inner color=mycolor11,outer color=mycolor11, fill opacity=0.2] (2.45,-2.45) circle (1cm);
\node [fill=black, shape=circle, inner sep=.4pt] ($.$) at (2.81,2.81) {};
\draw (2.81,2.81) circle (1cm);
\draw [inner color=mycolor11,outer color=mycolor11, fill opacity=0.2] (2.81,2.81) circle (1cm);
\node [fill=black, shape=circle, inner sep=.4pt] ($.$) at (-2.45,2.45) {};
\draw (-2.45,2.45) circle (1cm);
\draw [inner color=mycolor11,outer color=mycolor11, fill opacity=0.2] (-2.45,2.45) circle (1cm);
\node [fill=black, shape=circle, inner sep=.4pt] ($.$) at (-2.81,-2.81) {};
\draw (-2.81,-2.81) circle (1cm);
\draw [inner color=mycolor11,outer color=mycolor11, fill opacity=0.2] (-2.81,-2.81) circle (1cm);
% \draw [top color=cyan!40, bottom color=white, fill opacity=0.2] (-2.81,-2.81) circle (1cm);

% \filldraw[inner color=cyan!40!gray,outer color=yellow!20] (0,0) circle (1.8);
%

\foreach \s in {3}
{
\draw [thick] (-\s,-\s) -- (\s,-\s) -- (\s,\s) -- (-\s,\s) -- (-\s,-\s);
}

\node [fill=black, shape=circle, inner sep=.4pt] ($.$) at (0,0) {};
% \draw[inner color=cyan!40!gray,outer color=yellow!5]
% --- Arrow
\draw [dashed] (0,0) -- (3,0) node [right,font=\small] {$A/2$};
\draw [dashed] (-2.81,-2.81) -- (-3.71,-2.83) node [left,font=\small] {$\sqrt{n\epsilon_n}$};
\draw [dashed] (3,-3) -- (-3,3) node [above left,font=\small] {$A\sqrt{n}$};
\end{tikzpicture}}
	\caption{Illustration of a saturated sphere packing inside a cube, where small spheres of radius $r_0 = \sqrt{n\epsilon_n}$ cover a larger cube. Yellow colored spheres are not entirely contained within the larger cube, and yet they contribute to the packing arrangement. As we assign a codeword to each sphere center, the $1$-norm and arithmetic mean of a codeword are bounded by $A$ as required.}
	\label{Fig.Density}
\end{figure}
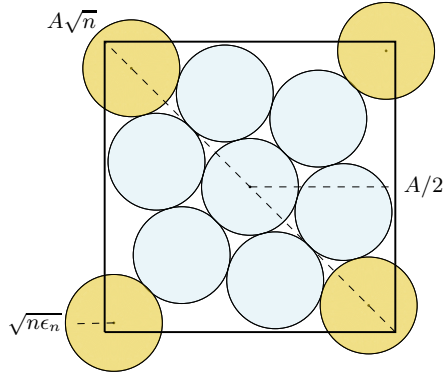

Let $\mathscr{S}$ denote a sphere packing, i.e., an arrangement of $L$ non-overlapping spheres $\S_{\fu_i}(n,r_0)$, $i\in [\![L]\!]$, that are packed inside the larger cube $\Q_{\f0}(n,A)$ with an edge length $A$, see Figure~\ref{Fig.Density}. As opposed to standard sphere packing coding techniques \cite{CHSN13}, the spheres are not necessarily entirely contained within the cube. That is, we only require that the centers of the spheres are inside $\Q_{\f0}(n,A)$ and are disjoint from each other and have a non-empty intersection with $\Q_{\f0}(n,A)$. The packing density $\Delta_n(\mathscr{S})$ is defined as the fraction of the cube volume $\text{Vol}\left[\Q_{\f0}(n,A)\right]$ that is covered by the spheres (see \cite[Ch.~1]{CHSN13}), i.e.,
%%%
\begin{align}
    \Delta_n(\mathscr{S}) \triangleq \frac{\text{Vol}\left(\Q_{\f0}(n,A)\cap\bigcup_{i=1}^{L}\S_{\fu_i}(n,r_0)\right)}{\text{Vol}\left[\Q_{\f0}(n,A)\right]} \,.\,
    \label{Eq.DensitySphereFast}
\end{align}
%%%
Sphere packing $\mathscr{S}$ is called \emph{saturated} if no spheres can be added to the arrangement without overlap.
% ---
% ---
In particular, we use a packing argument that has a similar flavor as that observed in the Minkowski--Hlawka theorem for saturated packing \cite{CHSN13}.
% We use the property that there exists an arrangement $\bigcup_{i=1}^{L} \S_{\fu_i}(n,\sqrt{n\epsilon_n})$ of non-overlapping spheres inside a $\Q_{\f0}(n,A)$, with a density of $\Delta_n(\mathscr{S})\geq 2^{-n}$ \cite[Lem.~2.1]{C10}.
Specifically, consider a saturated packing arrangement of 
% ---
\begin{align}
    \bigcup_{i=1}^{L(n,R)} \S_{\fu_i}(n,\sqrt{n\epsilon_n})
\end{align}
% ---
spheres with radius $r_0=\sqrt{n\epsilon_n}$ embedded within cube $\Q_{\f0}(n,A)$. Then, for such an arrangement, we have the following lower \cite[Lem.~2.1]{C10} and upper bounds \cite[Eq.~45]{CHSN13} on the packing density 
% ---
\begin{align}
    \label{Ineq.Density}
    2^{-n} \leq \Delta_n(\mathscr{S}) \leq 2^{-0.599n} \;.\,
\end{align}
% ---
The volume of a hyper sphere of radius $r$ is given by \cite[Eq.~(16)]{CHSN13}
% ---
\begin{align}
    \text{Vol}\left(\S_{\fx}(n,r)\right) = \frac{\pi^{\frac{n}{2}}}{\Gamma(\frac{n}{2}+1)} \cdot r^{n} \,.\,
    \label{Eq.VolS}
\end{align}
% ---
We assign a codeword to the center $\fu_i$ of each small sphere. The codewords satisfy the input constraint as $0 \leq u_{i,t} \leq A$, $\forall t \in [\![n]\!]$, $\forall i\in [\![L]\!]$, which is equivalent to
% ---
\begin{align}
    \label{Ineq.Norm_Infinity}
    \norm{\fu_i}_{\infty} \leq A \;.\,
\end{align}
% ---
Since the volume of each sphere is equal to $\text{Vol}(\S_{\fu_1}(n,r_0))$ and the centers of all spheres lie inside the cube, the total number of spheres is bounded from below by
%---
\begin{align}
    \label{Eq.L_Achiev}
    L & = \frac{\text{Vol}\left(\bigcup_{i=1}^{L}\S_{\fu_i}(n,r_0\right)}{\text{Vol}(\S_{\fu_1}(n,r_0))}
    \nonumber\\
    & \geq \frac{\text{Vol}\left(\Q_{\f0}(n,A)\cap\bigcup_{i=1}^{L}\S_{\fu_i}(n,r_0)\right)}{\text{Vol}(\S_{\fu_1}(n,r_0))}
    \nonumber\\
    & = \frac{\Delta(\mathscr{S}) \cdot
    \text{Vol}\left[\Q_{\f0}(n,A)\right]}{\text{Vol}(\S_{\fu_1}(n,r_0))}
    \nonumber\\
    & \geq 2^{-n} \cdot \frac{A^n}{\text{Vol}(\S_{\fu_1}(n,r_0))}
    % \nonumber\\
    % & = 2^{-n}\cdot \frac{A^n}{\text{Vol}(\S_{\fu_1}(n,\sqrt{n\epsilon_n}))}
    \,,\,
\end{align}
% ---
where the first inequality holds by (\ref{Eq.DensitySphereFast}) and the second inequality holds by (\ref{Ineq.Density}).
% ---
%\begin{align}
 %   \label{Eq.L}
 %   L \geq 2^{-n}\cdot %\frac{A^n}{\text{Vol}(\S_{\fu_1}(n,\sqrt{n\epsilon_n}))%} \,.\,
%\end{align}
% ---
The above bound can be further simplified as follows
% ---
\begin{align}
    \log L & \geq \log \left( \frac{A^n}{\text{Vol}\left(\S_{\fu_1}(n,r_0)\right)} \right) - n
    \nonumber\\
    & \stackrel{(a)}{=} n\log\left( \frac{A}{\sqrt{\pi} r_0 } \right)+\log \left(\frac{n}{2}! \right) - n
    \nonumber\\
    & \stackrel{(b)}{=} n \log A - n \log r_0 + \frac{1}{2} n \log n - n \log e + o(n) \;,\,
\end{align}
% ---
where $(a)$ exploits (\ref{Eq.VolS}) and $(b)$ follows from Stirling's approximation, see Appendix~\ref{App.SP_Diverging_Radius}. Now, for $r_0 = \sqrt{n\epsilon_n} = \sqrt{a}n^{\frac{1+b}{4}}$, we obtain
% ---
\begin{align}
    \log L & \geq
    n \log \frac{A}{\sqrt{a}} - \frac{1}{4}(1+b) \, n \log n + \frac{1}{2} n \log n - n \log e + o(n)
    \nonumber\\
    & = \left( \frac{1-b}{4} \right) \, n \log n + n ( \log \frac{A}{e\sqrt{a}}) + o(n)
    \;,\,
    \label{Eq.Log_L}
\end{align}
% ---
where the dominant term is of order $n \log n$. Hence, for obtaining a finite value for the lower bound of the rate, $R$, \eqref{Eq.Log_L} induces the scaling law of $L$ to be $2^{(n\log n)R}$. Therefore, we obtain
% ---
\begin{align}
    R & \geq \frac{1}{n\log n} \left[ \left( \frac{1-b}{4} \right) n\log n + n \log \left( \frac{A}{e\sqrt{a}} \right) + o(n) \right] \;,\,
\end{align}
% ---
which tends to $\frac{1}{4}$ when $n \to \infty$ and $b\rightarrow 0$.
% ---
\subsubsection*{Encoding}
Given message $i\in [\![L]\!]$, transmit $\fx=\fu_i$.
\subsubsection*{Decoding}
Let
% ---
\begin{align}
    \delta_n = c\rho^2\epsilon_n  = c\rho^2an^{\frac{1}{2}(b-1)} \;,\,
    % \frac{4\rho^2A}{3n^{\frac{1}{2}(1-b)}} \,.\,
    \label{Eq.Delta_n}
\end{align}
% ---
where $0 < b < 1$  is an arbitrarily small constant and $0<c<2$ is a constant. To identify whether message $j\in \M$ was sent, the decoder checks whether the channel output $\mathbf{y}$ belongs to the following decoding set:
% ---
\begin{align}
    \label{Eq.Decoding_Set}
    \D_j & = \left\{ \fy \in \Y^n \;:\, \left| D(\fy;\fu_j) \right| \leq \delta_n \right\} \;,\,
\end{align}
% ---
where
% ---
\begin{align}
    D(\fy;\fu_j) = \frac{1}{n} \sum_{t=1}^n \left[ \left( y_t - \left( \rho u_{j,t} + \lambda \right) \right)^2 - \left( \rho u_{j,t} + \lambda \right) \right]
\end{align}
% ---
is referred to as the decoding metric evaluated for observation vector $\fy$ and codeword $\fu_j$.

\subsubsection*{Error Analysis}
Consider the type I errors, i.e., the transmitter sends $\fu_i$, yet $\fY\notin\D_i$. For every $i\in[\![L]\!]$, the type I error probability is bounded by
% ---Let $Y_t(i) \sim \text{Pois}( \lambda + \rho u_{i,t})$ denote the channel output at time $t$ given that $\fx=\fu_i$. 
% ---
\begin{align}
    P_{e,1}(i) & = \Pr\left( \left| D\left(\fY;\fu_i\right) \right| > \delta_n \, \big| \, \fu_i \right) \;,\, 
    \label{Eq.TypeIError}
\end{align}
% ---
where the condition means that $\fx = \fu_i$ was sent. In order to bound $P_{e,1}(i)$, we apply Chebyshev's inequality, namely
% ---
\begin{align}
    \label{Eq.Chebyshev}
    \Pr\left(\left| D\left( \fY; \fu_i \right) - \mathbb{E} \left\{ D\left(\fY; \fu_i\right) \, \big| \, \fu_i \right\} \right| > \delta_n \big| \fu_i \right) \leq \frac{\text{var}\left\{ D\left(\fY;\fu_i \right) \, \big| \, \fu_i \right\}}{\delta_n^2} \;.\,
    % \Pr(|D(\fY(i))| > \delta_n) \leq \frac{\mathbb{E}\{|D(\fY(i))|^2\}}{\delta_n^2} \;.\,
\end{align}
% ---
First, we derive the expectation of the decoding metric as follows
% ---
\begin{align}
     \mathbb{E} \left\{ D(\fY; \fu_i) \, | \, \fu_i \right\} & = \frac{1}{n} \sum_{t=1}^n \left[ \text{var} \left\{ Y_t  \, | \, u_{i,t} \right\} - \left( \rho u_{i,t} + \lambda \right)  \right]  = 0 \,.\,
    % \mathbb{E} \left\{ D(\fY(j)) \right\}
    % & = \frac{1}{n} \sum_{t=1}^n \left[ \text{var} \left\{ Y_t(i) \right\} - \left( \rho u_{i,t} + \lambda \right) \right]  = 0 \,.\,
    \label{Eq.Expectation}
\end{align}
% ---
% Hence, the expectation of the decoding variable
%the random variable within the absolute value in (\ref{Eq.TypeIError}) 
% is zero.
%
Now, since the channel is memoryless, we can compute the variance as follows
% ---
\begin{align}
    \text{var} \left\{ D(\fY;\fu_i ) \, | \, \fu_i \right\} = \frac{1}{n^2} \sum_{t=1}^n \text{var} \left\{ \left( Y_t - \left( \rho u_{i,t} + \lambda \right) \right)^2  \, \Big| \, u_{i,t} \right\} \;.\,
\end{align}
% ---

Next, we present a useful lemma.
% ---
\begin{lemma}
\label{Lem.MGF}
Let $Z\sim\text{Pois}(\lambda_Z)$ be a Poisson RV with mean $\lambda_Z$. The following inequality holds
\begin{align*}
    \mathbb{E} \left\{ \left( Z - \lambda_Z \right)^4 \right\} \leq 7
    \left( \lambda_Z^4 + \lambda_Z^3 + \lambda_Z^2 + \lambda_Z \right) \,.\,
\end{align*}
\end{lemma}
% ---
\begin{IEEEproof}
The proof is provided in Appendix~\ref{App.MGF}.
\end{IEEEproof}
 
Using the above lemma, we bound the variance of the decoding metric as follows
% ---
\begin{align}
    \mathbb{E} \left\{ \left| D(\fY;\fu_i) \right|^2 \, \big| \, \fu_i\right\} & \stackrel{(a)}{=}
    \text{var} \left\{ D(\fY;\fu_i) \, \big| \, \fu_i \right\} 
    \nonumber\\
    & \stackrel{(b)}{\leq} \mathbb{E} \left\{ \left| D(\fY;\fu_i) \right|^2 \, \big| \, \fu_i \right\}
    \nonumber\\
    & \stackrel{(c)}{=} \frac{1}{n} \mathbb{E} \left[ \left( Y_t - \left( \rho u_{i,t} + \lambda \right) \right)^4 \, \Big| \, u_{i,t} \right] \nonumber\\
    & \leq \frac{7}{n} \left( \left( \rho A + \lambda \right)^4 + \left( \rho A + \lambda \right)^3 + \left( \rho A + \lambda \right)^2 + \left( \rho A + \lambda \right) \right)
    \,.\,
    \label{Ineq.4thMoment}
\end{align}
% ---
where $(a)$ follows since $\mathbb{E}\left\{ D(\fY;\fu_i) \right\} = 0$, $(b)$ follows since $\text{var}\{Z\} \leq \mathbb{E}\left\{Z^2\right\}$, and $(c)$ holds by letting $Z = \left( Y_t - \left( \rho u_{i,t} + \lambda \right) \right)^2 $ and exploiting an upper bound on the fourth non-central moment of a Poisson random variable (see Appendix~\ref{App.MGF}). Therefore, exploiting \eqref{Eq.Chebyshev}, (\ref{Eq.Expectation}) and (\ref{Ineq.4thMoment}), we can bound the type I error probability in (\ref{Eq.TypeIError}) as follows
% ---
\begin{align}
    P_{e,1}(i) &= \Pr \left( \left| D(\fY;\fu_i) \right| > \delta_n \, \big| \, \fu_i \right)
    \label{Ineq.TypeI}
     \nonumber\\
     & \leq \frac{7 \left( \left( \rho A + \lambda \right)^4 + \left( \rho A + \lambda \right)^3 + \left( \rho A + \lambda \right)^2 + \left( \rho A + \lambda\right) \right)}{n\delta_n^2}
    % \nonumber\\
    %  & = \frac{3\lambda^2+\lambda}{n(\frac{A}{3n^{\frac{1}{2}(1-b)}})^2}
    \nonumber\\
     & = \frac{7 \left( \left( \rho A + \lambda \right)^4 + \left( \rho A + \lambda \right)^3 + \left( \rho A + \lambda \right)^2 + \left( \rho A + \lambda \right) \right)}{c^2\rho^4 a^2 n^b}
    \nonumber\\&
     \leq \lambda_1 \,,\,
\end{align}
% ---
for sufficiently large $n$ and arbitrarily small $\lambda_1>0$.
% , where the first inequality follows by Chebyshev's inequality.

Next, we address type II errors, i.e., when $\fY\in\D_j$ while the transmitter sent $\fu_i$.
Then, for every $i,j\in[\![L]\!]$, where $i\neq j$, the type II error probability is given by
% ---
\begin{align}
    P_{e,2}(i,j) = \Pr \left( \left| D(\fY;\fu_j) \right| \leq \delta_n \, \big| \, \fu_i \right) \;.\,
    \label{Eq.Pe2G}
\end{align}
% ---
Then, we have
% ---
\begin{align}
    D(\fY;\fu_j) = \underset{B}{\underbrace{\frac{1}{n} \sum_{t=1}^n \left( Y_t - \left( \rho u_{i,t} + \lambda \right) + \rho \left( u_{i,t} - u_{j,t} \right) \right)^2}} - \frac{1}{n} \sum_{t=1}^n \left( \rho u_{j,t} + \lambda \right)  \;.\,
\end{align}
% ---
Observe that term $B$ can be expressed as follows
% ---
\begin{align}
    B = \frac{1}{n}\left[\norm{ \fY - \left( \rho \fu_i + \lambda \boldsymbol{1}_n \right)}^2 + \norm{\rho \left( \fu_i - \fu_j \right)}^2 +  2\rho \sum_{t=1}^n \left( u_{i,t} - u_{j,t} \right) \left( Y_t - \left( \rho u_{i,t} + \lambda \right) \right)\right] \;.\,
     \label{Eq.Pe2norm}
\end{align}
% ---
Then, define the  following events
% ---
\begin{align*}
    \H_i^j & = \left\{ \left| \frac{1}{n} \sum_{t=1}^n \left[ \left( Y_t - \left( \rho u_{i,t} + \lambda \right) + \rho  \left( u_{i,t} - u_{j,t} \right) \right)^2 - \left( \rho u_{j,t} + \lambda \right) \right] \right| \leq \delta_n \, \Big| \, u_{i,t}  \right\} \\
    \E_0 &= \left\{\left|\frac{2\rho}{n} \sum_{t=1}^n \left( u_{i,t} - u_{j,t} \right) \left( Y_t - \left(  {\rho} u_{i,t} + \lambda \right) \right) \right| > \delta_n \, \Big| \, \fu_i \right\} \\
    \E_1
    & = \left\{ \frac{1}{n}\left[\norm{\fY - ({ \rho} \fu_i + \lambda\boldsymbol{1}_n)}^2 + \big\|\rho\left(\fu_i-\fu_j\right)\big\|^2 - \sum_{t=1}^n \left( \rho u_{j,t} + \lambda \right)\right] \leq 2\delta_n  \, \Big| \, \fu_i \right\} \,.\,
\end{align*}
% ---
Exploiting the reverse triangle inequality, i.e., $\left| \beta \right| - \left| \alpha \right| \leq \left| \beta - \alpha \right|$, and letting $\beta = B$ and $\alpha = \frac{1}{n} \sum_{t=1}^n \left( \rho u_{j,t} + \lambda \right)$, we obtain the following upper bound on the type II error probability
% ---
\begin{align}
    P_{e,2}(i,j)  = \Pr\left( \H_i^j \right)
    %\nonumber\\
    & = \Pr\left( \left| \beta - \alpha \right| \leq \delta_n \right)
    \nonumber\\
    & \leq \Pr\left( \left| \beta \right| - \left| \alpha \right| \leq \delta_n \right)
    \nonumber\\
    % & = \Pr\left( |\beta| \leq \delta_n + |\alpha| \right)
    % \nonumber\\
    & = \Pr\left( |B| - \frac{1}{n} \sum_{t=1}^n \left( \rho u_{j,t} + \lambda \right) \leq \delta_n  \right) 
    \nonumber\\
    & \leq \Pr\left( B - \frac{1}{n} \sum_{t=1}^n \left( \rho u_{j,t} + \lambda \right) \leq \delta_n  \right) \;.\,
\end{align}
% ---
Now, applying the law of total probability to event
% ---
\begin{align}
    \B = \left\{ B - \frac{1}{n} \sum_{t=1}^n \left( \rho u_{j,t} + \lambda \right) \leq \delta_n  \right\} \,,\,
\end{align}
% ---
over $ \E_0$ and its complement $\E_0^c$, we obtain
% ---
\begin{align}
    P_{e,2}(i,j) & \leq \Pr \left( \B \cap \E_0 \right) + \Pr \left( \B \cap \E_0^c \right)
    \nonumber\\
    & \leq \Pr\left(\E_0 \right) + \Pr\left(\E_1 \right) \,,\;
    \label{Eq.Pe2_P_Expanded}
\end{align}
% ---
where the first expression in the last line follows since $\H_i^j \cap \E_0 \subset \E_0$. The latter expression holds since, given events $\E_0^c$ and $\B$, we immediately get event $\E_1$.

We now proceed with bounding $\Pr\left(\E_0 \right)$. By Chebyshev's inequality, the probability of this event can be bounded as follows
%%%
\begin{align}
    \Pr(\E_0) & \leq \frac{\text{var}\Big\{ \sum_{t=1}^n \left( u_{i,t} - u_{j,t} \right) \big( Y_t - \left( \rho u_{i,t} + \lambda \right) \big) \, \big| \, \fu_i \Big\}}{{n^2\delta_n^2}/{(4\rho^2)}} \nonumber\\
    & = \frac{4 {\rho^2} \sum_{t=1}^n (u_{i,t}-u_{j,t})^2\cdot\text{var}\{Y_t \, | \, u_{i,t}
    \}}{n^2\delta_n^2}
    \nonumber\\
    & = \frac{4{\rho^2} \sum_{t=1}^n(u_{i,t}-u_{j,t})^2\cdot({\rho} u_{i,t}+\lambda)}{n^2\delta_n^2}
    \nonumber\\
    & \leq \frac{4{\rho^2} ({\rho} A+\lambda)\sum_{t=1}^n(u_{i,t}-u_{j,t})^2}{n^2\delta_n^2} \nonumber\\
    & = \frac{4{\rho^2} ({\rho} A+\lambda)\norm{\fu_i-\fu_j}^2}{n^2\delta_n^2} \,.\,
    \label{Eq.PeE0G}
\end{align}
% --- 
Observe that
% ---
\begin{align}
    \norm{\fu_i - \fu_j}^2 & \stackrel{(a)}{\leq} \left(\norm{\fu_i} + \norm{\fu_j}\right)^2 \nonumber\\
    & \stackrel{(b)}{\leq} \left(\sqrt{n} \norm{\fu_i}_{\infty} + \sqrt{n} \norm{\fu_j}_{\infty} \right)^2
    \nonumber\\
    & \stackrel{(c)}{\leq} \left(\sqrt{n} A + \sqrt{n} A \right)^2
    \nonumber\\
    & = 4nA^2 \;,\
\end{align}
% ---
where $(a)$ holds by the triangle inequality, $(b)$ follows since $\norm{\cdot} \leq \sqrt{n} \norm{\cdot}_{\infty}$, and $(c)$ is valid by (\ref{Ineq.Norm_Infinity}). Hence, we obtain
% ---
\begin{align}
    \label{Ineq.E_0}
    \Pr(\E_0) & \leq \frac{4{\rho^2} ({\rho} A+\lambda)\norm{\fu_i-\fu_j}^2}{n^2\delta_n^2}
    \nonumber\\
    & \leq \frac{16 n {\rho^2} ({\rho} A+\lambda) A^2}{n^2\delta_n^2}
    \nonumber\\
    & = \frac{16 {\rho^2} ({\rho} A+\lambda) A^2}{n\delta_n^2}
    \nonumber\\
    & = \frac{16 ({\rho} A+\lambda) A^2}{c^2\rho^2 a^2 n^b}
    %\nonumber\\&
    % & = \frac{16 ({\rho} A+\lambda) }{{\rho} n^b}
    \nonumber\\
    & \leq \zeta_0 \,,\,
\end{align}
% ---
for sufficiently large $n$, where $\zeta_0 > 0$ is an arbitrarily small constant.
% Furthermore, since the event $\E_0^c$ happens with probability one, we focus on result of $\E_0^c$, namely
% % ---
% \begin{align}
%     \label{Eq.E_0_Result}
%     2\rho \sum_{t=1}^n \left( u_{i,t} - u_{j,t} \right) \left( Y_t - \left( {\rho} u_{i,t} + \lambda  \right) \right) \geq -n\delta_n \,.\,
% \end{align}
% % ---

We now proceed with bounding $\Pr\left(\E_1 \right)$ as follows. Based on the codebook construction, each codeword is surrounded by a sphere of radius $\sqrt{n\epsilon_n}$, that is 
% ---
 \begin{align}
    \label{Eq.distanceU}
     \norm{\fu_i - \fu_j}^2 \geq 4n\epsilon_n \;.\,
 \end{align}
% ---
Thus, we can establish the following upper bound for event $\E_1$:
% ---
\begin{align}
    \label{Ineq.E_1}
    \Pr(\E_1) & = \Pr\left( \frac{1}{n}\left[\norm{\fY - ({ \rho} \fu_i + \lambda\boldsymbol{1}_n)}^2 + \big\|\rho\left(\fu_i-\fu_j\right)\big\|^2 - \sum_{t=1}^n \left( \rho u_{j,t} + \lambda \right) \right] \leq 2\delta_n \, \Big| \, \fu_i \right)
    % \nonumber\\
    % & \leq \Pr\left( \frac{1}{n}\norm{\fY - \left( { \rho} \fu_i + \lambda\boldsymbol{1}_n \right)}^2 \leq 2\delta_n - \rho^2 \cdot \norm{\fu_i - \fu_j}^2 +\frac{1}{n} \sum_{t=1}^n \left( \rho u_{j,t} + \lambda \right) \, \Big| \, \fu_i \right)
    \nonumber\\
    & \stackrel{(a)}{\leq} \Pr\left( \frac{1}{n} \left[ \norm{\fY - \left( { \rho} \fu_i + \lambda\boldsymbol{1}_n \right)}^2 - \sum_{t=1}^n \left( \rho u_{j,t} + \lambda \right) \right] \leq 2(c-2)\rho^2\epsilon_n \; \, \Big| \, \fu_i \right)
    % \nonumber\\
    % & = \Pr\left( \frac{1}{n} \norm{\fY - \left( { \rho} \fu_i + \lambda\boldsymbol{1}_n \right)}^2 - \left( \frac{1}{n} \sum_{t=1}^n \left( \rho u_{j,t} + \lambda \right) \right) \leq - \frac{4n\epsilon_n\rho^2}{3} \, \Big| \, \fu_i \right)
    \nonumber\\
    & = \Pr\left( \frac{1}{n} \left[ \sum_{t=1}^n \left( Y_t - \left( \rho u_{i,t} + \lambda \right) \right)^2 - \left( \rho u_{j,t} + \lambda \right) \right] \leq 2(c-2)\rho^2\epsilon_n \; \, \Big| \, \fu_i \right)
    \nonumber\\
    & \stackrel{(b)}{\leq} \frac{\text{var}\left\{ \frac{1}{n} \sum_{t=1}^n \left( Y_t - \left( \rho u_{i,t} + \lambda \right) \right)^2 \right\}}{\left( 2(c-2)\rho^2\epsilon_n \right)^2}
    \nonumber\\
    & \stackrel{(c)}{\leq} \frac{7 \left( \left( \rho A + \lambda \right)^4 + \left( \rho A + \lambda \right)^3 + \left( \rho A + \lambda \right)^2 + \left( \rho A + \lambda \right) \right)}{n\left( 2(c-2)\rho^2\epsilon_n \right)^2}
    \nonumber\\
    & = \frac{7 \left( \left( \rho A + \lambda \right)^4 + \left( \rho A + \lambda \right)^3 + \left( \rho A + \lambda \right)^2 + \left( \rho A + \lambda \right) \right)}{ 4(c-2)^2 \rho^4 a^2 n^b }
    \nonumber\\
    & \leq \zeta_1 \;,\,
    % \nonumber\\
    % & = \frac{7 \left( \left( \rho A + \lambda \right)^4 + \left( \rho A + \lambda \right)^3 + \left( \rho A + \lambda \right)^2 + \left( \rho A + \lambda \right) \right)}{ 16 a \rho^4 n^b }
 \end{align}
% ---
for sufficiently large $n$, where $\zeta_1 > 0$ is an arbitrarily small constant. Here, $(a)$ follows from \eqref{Eq.distanceU} and (\ref{Eq.Delta_n}), $(b)$ holds by Chebyshev's inequality as given in (\ref{Eq.Chebyshev}), and $(c)$ follows by Lemma~\ref{Lem.MGF}. Therefore,
% ---
\begin{align}
    P_{e,2}(i,j) & \leq \Pr(\E_0) + \Pr(\E_1) 
    \nonumber\\
    & \leq \zeta_0 + \zeta_1 
    \nonumber\\
    & \leq \lambda_2 \,,\,
\end{align}
% ---
We have thus shown that for every $\lambda_1,\lambda_2>0$ and sufficiently large $n$, there exists an $(L(n,R), n, \allowbreak \lambda_1, \lambda_2)$ code.
% -------------------------
\subsection{Converse Proof}
\label{Subsec.ConvFast}
We show that the capacity is bounded by $\mathbb{C}_{DI}(\W,L)\leq \frac{3}{2}$. The derivation of this upper bound for the achievable rate of the DTPC is more involved than the derivation in the Gaussian case \cite{Salariseddigh_ITW}. In our previous work on Gaussian channels with fading \cite{Salariseddigh_ITW}, the converse proof was based on establishing a minimum distance between each pair of codewords. Here, on the other hand, we use the stronger requirement that the ratio of the letters of every two different codewords is different from $1$ for at least one index.

%  Given a codebook $\U^{(n)}=\{\fu_i\}$, for $i\in [\![L]\!]$ and all $t \in [\![ n ]\!]$ and we define the shifted codewords $\fv_i$ by
% % ---
% \begin{align*}
%     v_{i,t} = \lambda + u_{i,t} \;.\,
% \end{align*}
% % ---
We begin with the following lemma on the ratio of the letters of every pair of codewords.
%%%
\begin{lemma}
\label{Lem.DConverseFast}
Suppose that $R$ is an achievable rate for the DTPC. Consider a sequence of $(L(n,R), n,\allowbreak \lambda_1^{(n)}, \allowbreak \lambda_2^{(n)})$ codes $(\U^{(n)},\D^{(n)})$ such that $\lambda_1^{(n)}$ and $\lambda_2^{(n)}$ tend to zero as $n\rightarrow\infty$. Then, given a sufficiently large $n$, the codebook 
$\U^{(n)}$ satisfies the following property.
For every pair of codewords, $\fu_{i_1}$ and $\fu_{i_2}$, there exists at least one letter $t \in [\![n]\!]$ such that 
% the ratio $\frac{v_{i_2,t}}{v_{i_1,t}}$ for corresponding letters of the two codewords is distanced from $1$ by at least $\epsilon'_n$. That is,
% ---
\begin{align}
    \label{Eq.Converse_Lem}
    \left|1-\frac{\rho u_{i_2,t}+\lambda}{\rho u_{i_1,t}+\lambda}\right| > \epsilon'_n \,,\,
\end{align}
% --------------------
for all $i_1,i_2\in [\![L]\!]$, such that $i_1\neq i_2$,
with
% ---
\begin{align}
\label{Eq.epsilonn_p}
  \epsilon'_n = \frac{P_{\,\text{max}}}{n^{1+b}} \,,\,
\end{align}
% ---
where $b>0$ is an arbitrarily small constant.
\end{lemma}
% ---
\begin{IEEEproof}
The proof is given in Appendix~\ref{App.Converse}.
\end{IEEEproof}
% ---
 Next, we use Lemma~\ref{Lem.DConverseFast} to prove the upper bound on the DI capacity. Observe that since
% ---
\begin{align*}
    \label{Ineq.Shifted_Codeword}
    v_{i,t} & = \rho u_{i,t} + \lambda > \lambda \;,\,
\end{align*}
% ---
Lemma~\ref{Lem.DConverseFast} implies
% ---
\begin{align}
     \rho \left| u_{i_1,t} - u_{i_2,t} \right| & = \left| v_{i_1,t} - v_{i_2,t} \right|
     \nonumber\\
     & \stackrel{(a)}{>} \epsilon'_n v_{i_1,t}
     \nonumber\\
     & \stackrel{(b)}{>} \lambda \epsilon'_n \;,\,
\end{align}
% ---
where $(a)$ follows by (\ref{Eq.Converse_Lem}) and $(b)$ holds by (\ref{Ineq.Shifted_Codeword}). Now, since $\norm{\fu_{i_1} - \fu_{i_2}} \geq \left| u_{i_1,t} - u_{i_2,t} \right|$, we deduce that the distance between every pair of codewords satisfies
% ---
\begin{align}
   \norm{\fu_{i_1} - \fu_{i_2}} > \frac{\lambda \epsilon'_n}{\rho} \;.\,
\end{align}
% ---
Thus, we can define an arrangement of non-overlapping spheres $\S_{\fu_i}(n,\lambda \epsilon'_n)$, i.e., spheres of radius $\lambda \epsilon'_n$ that are centered at the codewords $\fu_i$. Since the codewords all belong to a hyper cube $\Q_{\f0}(n,P_{\,\text{max}})$ with edge length $P_{\,\text{max}}$, it follows that the number of packed small spheres, i.e., the number of codewords $L$, is bounded by
% ---
\begin{align}
    \label{Eq.L}
    L & = \frac{\text{Vol}\left(\bigcup_{i=1}^{L}\S_{\fu_i}(n,r_0\right)}{\text{Vol}(\S_{\fu_1}(n,r_0))}
    \nonumber\\
    & \leq \frac{\Delta(\mathscr{S}) \cdot
    \text{Vol}\left[\Q_{\f0}(n,P_{\,\text{max}})\right]}{\text{Vol}(\S_{\fu_1}(n,r_0))}
    \nonumber\\
    & \leq 2^{-0.599n} \cdot\frac{P_{\,\text{max}}^n}{\text{Vol}(\S_{\fu_1}(n,r_0))} \;,\,
\end{align}
% ---
where the last inequality follows from inequality (\ref{Ineq.Density}). Thereby,
% ---
\begin{align}
    \label{Eq.Converse_Log_L}
    \log L & \leq \log \left( \frac{P_{\,\text{max}}^n}{\text{Vol}\left(\S_{\fu_1}(n,r_0)\right)} \right) - 0.599n
    \nonumber\\
    % & = n\log\left( \frac{P_{\,\text{max}}}{\sqrt{\pi} r_0 } \right)+\log \left(\frac{n}{2}! \right) - 0.599n
    % \nonumber\\
    % & = n \log P_{\,\text{max}} - n \log r_0 + \frac{1}{2} n \log n - n \log(e) + o(n) \;,\,
    & = n \log P_{\,\text{max}} - n \log r_0 - n \log \sqrt{\pi} + \frac{1}{2}n\log \frac{n}{2} - \frac{n}{2}\log e + o(n) - 0.599n \,,\;
\end{align}
% ---
where the dominant term is again of order $n \log n$. Hence, for obtaining a finite value for the upper bound of the rate, $R$, \eqref{Eq.Converse_Log_L} induces the scaling law of $L$ to be $2^{(n\log n)R}$. Hence, by setting $L(n,R) = 2^{(n\log n)R}$ and $r_0 = \frac{\lambda \epsilon'_n}{2\rho} = \frac{\lambda P_{\,\text{max}}}{2\rho n^{1+b}}$, we obtain
% ---
\begin{align}
    R & \leq \frac{1}{n\log n} \left[ n \log P_{\,\text{max}} - n \log r_0 - n \log \sqrt{\pi} + \frac{1}{2}n\log \frac{n}{2} - \frac{n}{2}\log e + o(n) - 0.599n \right]
    \nonumber\\
    & = \frac{1}{n\log n} \left[ \left( \frac{1}{2} + \left( 1 + b \right) \right) \, n \log n - n \left( \log \frac{\lambda\sqrt{\pi e}}{2\rho} + 1.0599 \right)+ o(n) \right]
    \;,\,
\end{align}
% ---
which tends to $\frac{3}{2}$ as $n \to \infty$ and $b \to 0$.  This completes the proof of Theorem~\ref{Th.PDICapacity}.

% Since the codewords all belong to a hyper cube $\Q_{\f0}(n,P_{\,\text{max}})$ with edge $P_{\,\text{max}}$, it follows that the number of packed small spheres, i.e., the number of codewords $2^{(n\log n)R}$, is bounded by
% %%%
% \begin{align}
%     2^{(n\log n)R} & \leq \frac{\text{Vol}\left[\Q_{\f0}(n,P_{\,\text{max}})\right]}{\text{Vol}(\S_{\fu_1}(n,\lambda \epsilon'_n))} = \frac{P_{\,\text{max}}^n}{\text{Vol}(\S_{\fu_1}(n,\lambda \epsilon'_n))}
% \end{align}
% %%%
% Now since $\log\text{Vol}\left(\S^n(\fu_1,r)\right) > -\frac{n}{2}\log(\frac{n}{2\pi e})-\frac{1}{2}\log n\pi - \frac{\log e}{6n} + n \log r$ \cite[See Eq.~16-18]{CHSN13}, and $ r = \lambda \epsilon'_n = \frac{\lambda P_{\,\text{max}}}{n^{1+b}}$, we obtain
% % ---
% \begin{align}
%     (n\log n)R & \leq n\log P_{\,\text{max}} +\frac{n}{2}\log(\frac{n}{2\pi e})+\frac{1}{2} \log n\pi + \frac{\log e}{6n} - n\log \lambda \epsilon'_n \,,\,
% \end{align}
% % ---
% then,
% % ---
% \begin{align}
%     R &\leq \frac{\log P_{\,\text{max}}}{\log n} + \frac{\log(\frac{n}{2\pi e})}{2\log n} + \frac{\log n\pi}{2n\log n} + \frac{\log e}{6n^2\log n} - \frac{\log \lambda \epsilon'_n}{\log n}
%     \nonumber\\
%     & \sim \frac{\log\left(n(2\pi e)^{-1}\right)}{2\log n} + \frac{\log\left(\lambda^{-1} P_{\,\text{max}}^{-1}n^{1+b}\right)}{\log n}
%     \nonumber\\
%     & \sim \frac{1}{2} + \frac{\log(n^{1+b})}{\log n}
%     \nonumber\\
%     & = \frac{1}{2} + (1+b) \cdot \frac{\log n}{\log n}
% \end{align}
% % ---
% which tends to $\frac{3}{2}$ as $n\to\infty$ and $b \to 0$.
% --------------------------
\section{Simulation Results}
\label{Sec.Simulation}
We emphasize that the main result of this paper is the characterization of the DI capacity for the DTPC (cf. Theorem~\ref{Th.PDICapacity}), which by definition holds for asymptotically large codewords, i.e., as $n\to\infty$. Nevertheless, the proposed achievability scheme presented in Section~\ref{sec:achievable} is based on a constructive proof, which allows us to generate a practical code even for finite $n$. Therefore, in the following, we evaluate the performance of an explicitly constructed codebook in terms of empirical type I and type II error rates.
%The simulation setup is as follows: We choose code rate to be $R = 0.1$. Expected number of interfering molecules is assumed to be $\lambda = 0.2$ and parameter $b$ (see \eqref{Eq:epsilon}) is set to $b = 0.99$. We assume codebook precision $\epsilon_n \in [9.834,9.853] = $ and decoding threshold $\delta_n \in [3.278,3.284]$. The power constraints $P_{\,\text{max}}$ and $P_{\,\text{ave}}$ are chosen to be some number greater or equal than $10$. Thus, the minimum value of them, i.e., that parameter $A$ is considered to be $A = 10$ in the program.
The values of the parameters used in the proposed simulation setup and codebook construction are summarized in Table~\ref{Table.I}. The codebook construction is briefly sketched in the following.  %Following the same line of argument provided in Subsection~\ref{Subsec.CodebookConstruction_1}, in order to observe the power constraints it suffices to observe only the condition for symbols of the codewords as follows: $0 \leq x_t \leq A$ $\forall \, t \in [\![n]\!]$ where $A$ is linked to the power constraints through \eqref{Eq.A}. To this aim, 
At first, codewords are generated uniformly, that is, the value of each symbol is chosen uniformly distributed between $0$ and $A$. Next, in order to realize the minimum distance property of the codebook, once a codeword is created, before adding it to the codebook, it is verified whether it has at least a minimum Euclidean distance of $2\sqrt{n\epsilon_n}$ from all previously generated codewords or not. In the course of codeword generation, if a codeword violates the minimum distance property, it is discarded and a new codeword is generated and the procedure is repeated until the desired codebook size is obtained. To simulate the receiver's task, the distance decoder in \eqref{Eq.Decoding_Set_0} is implemented and the empirical type I and type II error rates for finite codeword lengths are obtained via Monte Carlo simulation. 
%A set of parameters that seems practically rational are chosen for the simulation. In particular, 
In practical MC systems, very large codeword lengths might not be feasible due to restrictions on time, energy, etc. Therefore, we focus on a range of small codeword lengths, i.e., $19 \leq n \leq 28$. Moreover, since rates $R\geq \frac{1}{4}$ are achievable by the proposed scheme only as $n\to\infty$, we choose a smaller rate, i.e., $R=0.1$, for codebook generation for finite $n$. However, we study a codebook with super-exponential size in $n$, i.e., $L=2^{(n\log n)R}$, which is the
\begin{center}
    \footnotesize
    \ctable[
	caption = Parameters of The Simulations,
	captionskip = -2.5ex,
	label = Table.I,
    pos = t
	]{llr}{}
	{\FL \textbf{Description} & \textbf{Notation} & \textbf{Value}
	\ML Minimum of power constraints & $A=\min \left(P_{\,\text{ave}},P_{\,\text{max}} \right)$ & 1000 molecules/s
	\ML Release time & $T_{\rm rls}$ & 1 s
	\ML Prob. molecules reaching the receiver & $p_{\rm ch}$ & 0.01
	\ML Expected number of interfering molecules & $\lambda$ & 0.2
	\ML Code rate & $R$ & 0.1
	\ML Codeword length & $n$ & [19 - 28]
	\ML Codebook size & $L = 2^{(n\log n) R}$ & [268 - 11273]
	\ML Codebook parameters & $a,b,c$ & $10^5$, 0.99, $\frac{1}{3}$
	\ML Codebook precision & $\epsilon_n = an^{\frac{1}{2}(b-1)}$ & [9.853 - 9.834]
	\ML Decoding threshold & $\delta_n = c\rho^2 \epsilon_n$ & [3.284 - 3.278]
	\ML Codebook minimum distance & $2\sqrt{n\epsilon_n}$ & [27.36 - 33.18]
	\ML Number of iterations & - & $7 \times 10^5$
	\LL
	\vspace{-15mm}
	}
\end{center}
key claim of Theorem~\ref{Th.PDICapacity}. Without loss of generality, we assume that the transmitter sends message $i=1$ and denote the empirical type I and type II error rates (average and maximum) by $\bar{e}_{1}(i)$, and $\bar{e}_{2}^{\text{ave}}$, $\bar{e}_{2}^{\text{max}}$, respectively.

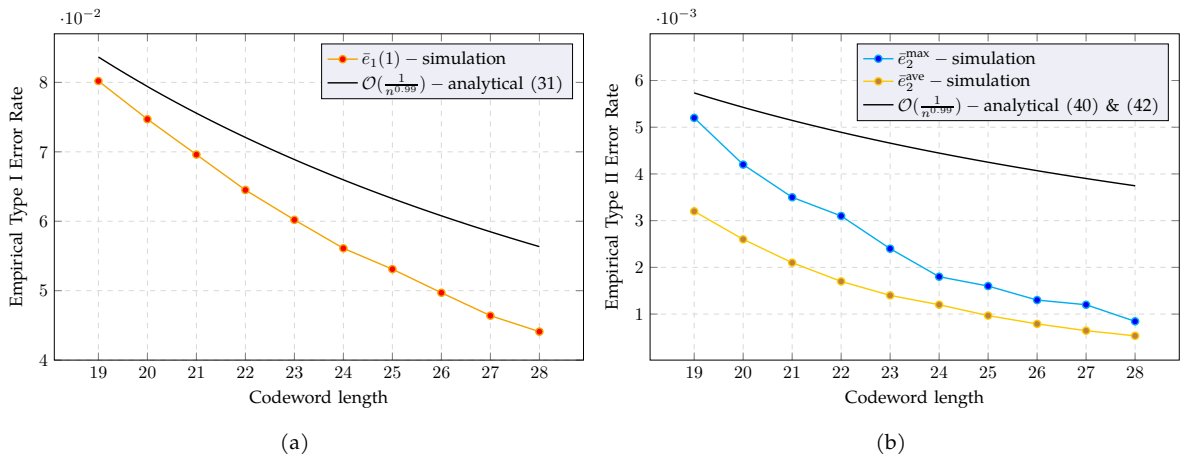
\begin{figure}[t]
    \centering
    \subfloat[][(a)]{\scalebox{.67}{
\begin{tikzpicture}[
    hplot/.style={ycomb, mark=square}]
    \begin{axis}[
    % extra y ticks = {0.0470},
    extra y tick labels = ,
    scale only axis,
    xticklabel style= {rotate=0,anchor=near xticklabel},
    legend cell align=left,
    legend columns={1},
    ymin = 0.04, ymax = .087,
    grid=both,
    ytick={.04,.05,.06,.07,.08},
    xlabel={Codeword length},
    ylabel={Empirical Type I Error Rate},
    legend style={at={(0.98,0.97)},anchor=north east},
    grid=major,
    grid style={dashed, gray!30},
    inner axis line style={-stealth},
    width=10.5cm, height=6.48cm,
    legend style={fill=mycolor4!10},
    legend cell align=left,
    yticklabel style={
    /pgf/number format/.cd,
    set decimal separator={.},
    precision=4,
    /tikz/.cd,},]
    \tikzmath{\c = ,.75;}
    \addplot[thick,orange_7b,mark=*,mark options={fill=red}]
    coordinates {
        (19,0.0802)
        (20,0.0747)
        (21,0.0696)
        (22,0.0645)
        (23,0.0602)
        (24,0.0561)
        (25,0.0531)
        (26,0.0497)
        (27,0.0464)
        (28,0.0441)
    };
\addlegendentry{$\bar{e}_{1}(1) - \text{simulation}$}
\addplot[thick,domain=19:28, yrange=0:.1, samples=1000] {1.58/(x^(0.99))-.002};
\addlegendentry{$\mathcal{O}(\frac{1}{n^{0.99}}) - \text{analytical \eqref{Ineq.TypeI}}$}
\end{axis}
\end{tikzpicture}}\label{Fig.Type_I}}
    \subfloat[][(b)]{\scalebox{.67}{
\begin{tikzpicture}[
    hplot/.style={ycomb, mark=square}]
    \begin{axis}[
    % extra y ticks = {0.00034286,0.00057,0.0008},
    extra y tick labels = ,
    scale only axis,
    xticklabel style= {rotate=0,anchor=near xticklabel},
    legend cell align=left,
    legend columns={1},
    ymin = 0.0000148, ymax = 0.007,
    grid=both,
    ytick={0.001,.002,.003,.004,.005,.006},
    xlabel={Codeword length},
    ylabel={Empirical Type II Error Rate},
    legend style={at={(0.98,0.97)},anchor=north east},
    grid=major,
    grid style={dashed, gray!30},
    inner axis line style={-stealth},
    width=10.5cm, height=6.48cm,
    legend style={fill=mycolor4!10},
    legend cell align=left,
    yticklabel style={
        /pgf/number format/.cd,
        set decimal separator={.},
        precision=4,
        /tikz/.cd,
        },
    ]
    \addplot [thick,cyan,mark=*,mark options={fill=blue}]
        coordinates {
        (19,0.0052)
        (20,0.0042)
        (21,0.0035)
        (22,0.0031)
        (23,0.0024)
        (24,0.0018)
        (25,0.0016)
        (26,0.0013)
        (27,0.0012)
        (28,0.00084480)
        };
    \addlegendentry{$\bar{e}_{2}^{\text{max}} - \text{simulation}$}
    
    \addplot [thick,orange_6b,mark=*,mark options={fill=brown}]
        coordinates {
        (19,0.0032)
        (20,0.0026)
        (21,0.0021)
        (22,0.0017)
        (23,0.0014)
        (24,0.0012)
        (25,0.00096724)
        (26,0.00079004)
        (27,0.00064463)
        (28,0.00053471)
        };
    \addlegendentry{$\bar{e}_{2}^{\text{ave}} - \text{simulation}$}
    
    \addplot[thick,domain=19:28, yrange=0:.01, samples=2000] {0.115/(x^(0.99)) - 0.0005};
    % \addplot[domain=19:22, yrange=0:.01, samples=2000] {500000*5/(e^x)};
    \addlegendentry{$\mathcal{O}(\frac{1}{n^{0.99}}) - \text{analytical \eqref{Ineq.E_0} \& \eqref{Ineq.E_1}}$}
\end{axis}

\end{tikzpicture}}\label{Fig.Type_II}}
    \caption{Impact of codeword length on the empirical type I and type II error rates. Larger lengths decrease the empirical rates.}
\end{figure}

In Figure~\ref{Fig.Type_I}, we observe that the empirical type I error rates decrease when the codeword length increases. Figure~\ref{Fig.Type_II} shows the empirical type II error rate, where a similar phenomenon as for the type I error rate is observed for the average and maximum error rates. The results in Figure~\ref{Fig.Type_I} and \ref{Fig.Type_II} suggest that any arbitrarily small values for the error probabilities can be achieved if the codeword length is sufficiently increased, which is consistent with the DI capacity result reported in Theorem~\ref{Th.PDICapacity}. These empirical observations are supported by the theoretical error analysis provided in the proof of Theorem~\ref{Th.PDICapacity}, where increasing the codeword length can reduce type I and II error probabilities below any arbitrarily small values. Furthermore, the simulation results in Figure~\ref{Fig.Type_I} and \ref{Fig.Type_II} show that the achieved error rates for the constructed code with $R = 0.1$, decay faster than the theoretical upper bounds provided in \eqref{Ineq.TypeI}, \eqref{Ineq.E_0} and \eqref{Ineq.E_1} evaluated for $b = 0.99$. Nevertheless, the general trend of the empirical error rates as functions of the codeword length is well captured by the analytical upper bounds.

We note that to fully characterize the asymptotic behavior of the decoding errors as a function of the codeword length for every value of the rate $0 < R < C$, knowledge of the corresponding channel reliability function is required \cite{Boche21}. To the best of the authors’ knowledge, the channel reliability function for DI has not been studied in the literature so far, neither for the Gaussian channel \cite{Salariseddigh_IT} nor the Poisson channel \cite{Salariseddigh_GC_IEEE,Salariseddigh_GC}. We note that even for the conventional message transmission problem, the characterization of the channel reliability function is difficult, as the corresponding channel reliability function is not Turing computable \cite{Boche21}.
% ---
% -------------------------------------
\section{Summary and Future Directions}
\label{Sec.Summary}

In this paper, we studied the DI problem over the DTPC, which may serve as a model for event-triggered based tasks in the context of MC for applications such as targeted drug delivery, health condition monitoring, olfactory systems, etc. In particular, we derived lower and upper bounds on the DI capacity of the DTPC subject to average and peak power constraints in the codebook size of $L(n,R)=2^{(n\log n)R}=n^{nR}$. Our results revealed that the super-exponential scale of $n^{nR}$ is the appropriate scale for the DI capacity of the DTPC, which was proved by finding a suitable sphere packing arrangement embedded in a hyper cube. We emphasize that this scale is sharply different from the ordinary scales in transmission and RI settings, where the codebook size grows exponentially and double exponentially, respectively.

The results presented in this paper can be extended in several directions, some of which are listed in the following as potential topics for future research works:
% ---
\begin{itemize}
    \item Our observations for the codebook size of the DTPC and Gaussian channels \cite{Salariseddigh_ITW} lead us to conjecture that the codebook size for any continuous alphabet channel would be a super-exponential function, i.e., $2^{(n\log n)R}$. However, a formal proof of this conjecture remains unknown.
    \item We assumed that the channel uses are orthogonal, which implies a memoryless channel for temporal codes and independent molecule reception for spatial codes. In practice, however, the DTPC may exhibit memory \cite{Gohari16} and non-orthogonal molecule reception \cite{Buck05}, the investigation of which constitutes an interesting research problem.
    \item This study has focused on a point-to-point system and may be extended to multi-user scenarios (e.g., broadcast and multiple access channels) or multiple-input multiple-output channels which become relevant in complex MC nano-networks.
    \item Another interesting research topic is to investigate the behavior of the DI capacity in the sense of Fekete's Lemma \cite{Boche20}, that is, to verify whether the pessimistic ($\underline{C} = \liminf_{n \to \infty} \allowbreak \frac{\log L(n,R)}{n \log n}$) and optimistic ($\overline{C} = \limsup_{n \to \infty} \frac{\log L(n,R)}{n \log n}$) capacities \cite{A06} are equal or not.
\end{itemize}
% ---
% ------------------------
\section*{Acknowledgments}
Salariseddigh was supported by the German Research Foundation (DFG) under grant DE 1915/2-1. Pereg and Deppe were supported by the German Federal Ministry of Education and Research (BMBF) under Grants 16KIS1005 (LNT, NEWCOM) and 16KISQ028. Boche was supported by the BMBF under Grant 16KIS1003K (LTI, NEWCOM), and the national initiative for ``Molecular Communications" (MAMOKO) under Grant 16KIS0914. Jamali was supported by the DFG under Grant JA 3104/1-1. Schober was supported by MAMOKO under Grant 16KIS0913.
% ---------
\appendices
\section{Volume of a Hyper Sphere With Growing Radius}
\label{App.SP_Diverging_Radius}
To solidify the idea of packing spheres within a hyper cube, we reveal and explain a counter-intuitive phenomenon regarding the packing of hyper spheres with growing radius in the codeword length inside a hyper cube. We observe that despite the fact that the hyper sphere's radius tends to infinity as the codeword length goes to infinity $\sim n^{\frac{1}{4}}$ its volume tends to zero. In fact, the volume of the sphere vanishes super-exponentially inverse, i.e., $\sim n^{-\frac{n}{4}}$, such that in the asymptotic analysis, we can accommodate a super-exponential number of such hyper spheres inside the hyper cube. We note that the ratio of the spheres in our construction grows with $n$, as $\sim n^{\frac{1}{4}}$.
It is well-known that the volume of an $n$-dimensional \emph{unit}-hyper sphere, i.e., with a radius of $r_0=1$, tends to zero, as $n\to\infty$ \cite[Ch.~1, Eq.~(18)]{CHSN13}. 
Nonetheless, we observe that the volume still tends to zero for a radius of $r_0 = n^c$, where  $0 < c < \frac{1}{2}$. More precisely,
% ---
\begin{align}
    \lim_{n\to\infty} \text{Vol}\left(\S_{\fu_1}(n,r_0)\right) & = \lim_{n\to\infty}\frac{\pi^{\frac{n}{2}}}{\Gamma(\frac{n}{2}+1)}\cdot r_0^n
    \nonumber\\&
     = \lim_{n\to\infty} \frac{\pi^{\frac{n}{2}}}{\frac{n}{2}!}\cdot r_0^n
    \nonumber\\&
     = \lim_{n\to\infty} \left(\sqrt{\frac{2\pi}{n}}r_0\right)^n \,,\,
    \label{Eq.nSphere_Volume}
\end{align}
% ---
where the last equality follows by Stirling's approximation \cite[P.~52]{F66}, that is,
% ---
\begin{align}
    \log n! = n \log n - n \log e + o(n) \;.\,
\end{align}
% ---
The last expression in (\ref{Eq.nSphere_Volume}) tends to zero for all $r_0 = n^c$ with $c \in (0,\frac{1}{2})$. On the other hand when $n \to \infty$, the volume of a hyper cube $\Q_{\f0}(n,A)$ with edge length $A$ is given by
% ---
\begin{align}
    \label{Eq.n_Cube_Vol}
    \lim_{n\to\infty} \text{Vol}\left[\Q_{\f0}(n,A)\right] = \lim_{n\to\infty} A^n = \begin{cases} 0 & A < 1 \,,\, \\ 1 & A = 1 \,,\, \\ \infty & A > 1 \,.\, \end{cases}
\end{align}
% ---
Now, to derive how many spheres can be packed inside the hyper cube $\Q_{\f0}(n,A)$ we derive the log-ratio of the volumes as follows
% ---
\begin{align}
    \label{Eq.Volume_Ratio}
    \log \left( \frac{\text{Vol}\left[\Q_{\f0}(n,A)\right]}{\text{Vol}\left(\S_{\fu_1}(n,r_0)\right)} \right)
    & = \log \left( \frac{A^n}{\pi^{\frac{n}{2}}{} r_0^n} \cdot \frac{n}{2}!\right)
    \nonumber\\
    & = n\log\left( \frac{A}{\sqrt{\pi} r_0 } \right)+\log \left(\frac{n}{2}! \right)
    \nonumber\\
    & = n \log A - n \log r_0  - n \log \sqrt{\pi} + \frac{1}{2}n\log \frac{n}{2} - \frac{n}{2}\log e + o(n)
    \nonumber\\
    % & = n \log A - n \log r_0 - n\log \sqrt{\pi} + \frac{1}{2} n \log n - n \left( \log \sqrt{e} + \frac{1}{2} \right) + o(n)
    % \nonumber\\
    % & \stackrel{(a)}{=} n \log A - n \log r_0 + \frac{1}{2} n \log n + n \left( \log A - \log \sqrt{e} - \frac{1}{2} \right) + o(n)
    % & = \left(\frac{1}{2} - c \right) \, n \log n + n \left( \log \frac{A}{\sqrt{\pi e}}  - \frac{1}{2} \right) + o(n)
    & = \left(\frac{1}{2} - c \right) \, n \log n + n \left( \log \frac{A}{\sqrt{\pi e}}  - \frac{3}{2} \right) + o(n)
    \;,\,
\end{align}
% ---
where the last equality follows from $r_0 = n^c$. Now, since the dominant term in (\ref{Eq.Volume_Ratio}) involves $n\log n$, we deduce that codebook size should be $L(n,R) = 2^{(n\log n)R}$, thereby by (\ref{Eq.L_Achiev}) we obtain
% ---
\begin{align}
    R & \geq \frac{1}{n\log n} \left[ \log \left( \frac{\text{Vol}\left[Q_{\f0}(n,A)\right]}{\text{Vol}\left(\S_{\fu_1}(n,r_0)\right)} \right) - n \right] 
    \nonumber\\
    & = \frac{1}{n\log n} \left[\left(\frac{1}{2} - c \right) \, n \log n + n \left( \log \frac{A}{\sqrt{\pi e}} - \frac{3}{2} \right) + o(n) \right] \,,\,
\end{align}
% ---
which tends to $\frac{1}{2} - c$ when $n \to \infty$.
% ----------------------------------------------------------
\section{Moment Generating Function of Poisson Random Variable}
\label{App.MGF}
The moment-generating function (MGF) of a Poisson variable $Z\sim\text{Pois}(\lambda_Z)$ is given by
% ---
\begin{align}
    G_Z(\alpha) = e^{\lambda_Z(e^{\alpha}-1)} \,.\,
\end{align}
% ---
Hence, for $X=Z-\lambda_Z$, the MGF is given by 
% ---
\begin{align}
    G_X(\alpha) = e^{\lambda_Z (e^{\alpha} - 1 - \alpha)} \,.\,
\end{align}
% ---
Since the fourth non-central moment equals the fourth order derivative of the MFG at $\alpha = 0$, we have
% ---
\begin{align*}
    \mathbb{E}\{ X^4 \} =
    \frac{d^4}{d\alpha^4}G_X(\alpha)
    \bigg|_{\alpha=0} &= \lambda_Z \left( \lambda_Z^3 e^{3\alpha} + 6\lambda_Z^2 e^{2\alpha} + 7\lambda_Z e^{\alpha} + 1 \right) e^{\alpha + \lambda_Z e^{\alpha} - \lambda_Z}
    \bigg|_{\alpha = 0} \nonumber\\
    & = \lambda_Z^4 + 6 \lambda_Z^3 + 7\lambda_Z^2 + \lambda_Z
    \nonumber\\&
     \leq 7
    \left( \lambda_Z^4 + \lambda_Z^3 + \lambda_Z^2 + \lambda_Z \right) \,.\,
\end{align*}
% ---

\section{Proof of Lemma~\ref{Lem.DConverseFast}}
\label{App.Converse}
In the following, we provide the proof of Lemma~\ref{Lem.DConverseFast}. The method of proof is by contradiction, namely, we assume that the condition given in \eqref{Eq.Converse_Lem} is violated and then we show that this leads to a contradiction (sum of the type I and type II error probabilities converge to one).

Fix $\lambda_1,\lambda_2 > 0$. Let $\kappa, \delta > 0$ be arbitrarily small constants. Assume to the contrary that there exist two messages $i_1$ and $i_2$, where $i_1\neq i_2$, meeting the error constraints in \eqref{Eq.ErrorConstraints}, such that for all $t\in[\![n]\!]$, we have
% ---
\begin{align}
    \label{Ineq.Converse_Lem_Complement}
    \left|1-\frac{v_{i_2,t}}{v_{i_1,t}}\right| \leq \epsilon'_n \;,\,
\end{align}
% ---
where $v_{i_k,t}=\rho u_{i_k,t}+\lambda,\,\,k=1,2$. In order to show contradiction, we will bound the sum of the two error probabilities, $P_{e,1}(i_1)+P_{e,2}(i_2,i_1)$, from below. To this end, define 
\begin{align}\label{Eq:Bi1}
       \B_{i_1} = \left\{\fy \in \D_{i_1} \,:\,
       \frac{1}{n}\sum_{t=1}^n y_t \leq  \rho P_{\,\text{max}} + \lambda  + \delta \right\} \;.\,
\end{align}
Then, observe that
% ---
    \begin{align}
    \label{Eq.Error_Sum_1}
    P_{e,1}(i_1)+P_{e,2}(i_2,i_1)
    & = 1- \sum_{\fy\in\D_{i_1}} W^n \left( \fy \, \big| \, \fu_{i_1} \right) + \sum_{\fy\in\D_{i_1}} W^n \left( \fy \, \big| \, \fu_{i_2} \right)
    \nonumber\\
    & \geq 1- \sum_{\fy\in\D_{i_1}} W^n \left( \fy \, \big| \, \fu_{i_1} \right) + \sum_{\fy\in\D_{i_1} \cap \B_{i_1}} W^n \left( \fy \, \big| \, \fu_{i_2} \right) \,.\,
    \end{align}
    % ---
    
    Now, consider the sum over $\D_{i_1}$ in (\ref{Eq.Error_Sum_1}),
    % ---
    \begin{align}
        \label{Ineq.Error_I_Complement}
        \sum_{\fy\in\D_{i_1}} W^n \left( \fy \, \big| \, \fu_{i_1} \right) & = \sum_{\fy\in\D_{i_1}\cap\B_{i_1}} W^n \left( \fy \, \big| \, \fu_{i_1} \right) + \sum_{\fy \in \D_{i_1}\cap\B_{i_1}^c} W^n \left( \fy \, \big| \, \fu_{i_1} \right)
        \nonumber\\
        & \leq \sum_{\fy \in \D_{i_1}\cap\B_{i_1}} W^n \left( \fy \, \big| \, \fu_{i_1} \right) + \Pr\left( \frac{1}{n} \sum_{t=1}^n Y_t >  \rho P_{\,\text{max}} + \lambda + \delta \, \bigg| \,  \fu_{i_1} \right) \,.\;
    \end{align}
    % ---
    Next, we bound the probability on the right hand side of (\ref{Ineq.Error_I_Complement}) as follows
    % ---
    \begin{align}
        & \Pr \left( \frac{1}{n} \sum_{t=1}^n Y_t - \frac{1}{n} \sum_{t=1}^n  \mathbb{E}\{Y_t\}    > \rho P_{\,\text{max}} + \delta - \frac{1}{n} \sum_{t=1}^n  \mathbb{E}\{Y_t\} \right)
        \nonumber\\
        & \overset{(a)}{\leq} \frac{\text{var} \left\{ \frac{1}{n} \sum_{t=1}^n Y_t \, \big| \, \fu_{i_1} \right\} }{\left( \rho P_{\,\text{max}} + \delta - \frac{1}{n} \sum_{t=1}^n  \mathbb{E}\{Y_t\} \right)^2} \nonumber\\
        & \overset{(b)}{=} \frac{ \frac{1}{n^2} \sum_{t=1}^n ( \rho u_{i_1,t} + \lambda) }{\left( \rho P_{\,\text{max}} + \delta - \frac{1}{n} \sum_{t=1}^n  \rho u_{i_1,t} + \lambda \right)^2}  \nonumber\\
        & \overset{(c)}{\leq} \frac{\rho P_{\max} + \lambda}{n\delta^2} 
        \nonumber\\&
         \leq \kappa
        \;,\,
        \label{Ineq.ErrorI_Complement1}
   \end{align}
% ---
for sufficiently large $n$, where inequality $(a)$ follows from Chebyshev's inequality, for equality $(b)$, we exploited $\text{var}\{Y_t|u_{i_1,t}\}=\mathbb{E}\{Y_t|u_{i_1,t}\}=\rho u_{i_1,t} + \lambda$, and for inequality $(c)$, we used the fact that $u_{i_1,t}\leq P_{\max} \,,\, \forall \, t\in[\![n]\!]$.

Returning to the sum of error probabilities in (\ref{Eq.Error_Sum_1}), exploiting the bound (\ref{Ineq.ErrorI_Complement1}) leads to
% ---
\begin{align}
    \label{Eq.Error_Sum_2}
    P_{e,1}(i_1)+P_{e,2}(i_2,i_1)
     & \geq 
    %  1- \Big[\sum_{\fy\in\D_{i_1}\cap\B_{i_1}} W^n(\fy|\fu_{i_1}) + \kappa \Big] + \sum_{\fy\in\D_{i_1}\cap\B_{i_1}}  W^n(\fy|\fu_{i_2})
    % % \nonumber\\
    % % & = 
    1 - \sum_{\fy \in \D_{i_1} \cap \B_{i_1}} \left[ W^n \left( \fy \, \big| \, \fu_{i_1} \right) - W^n \left( \fy \, \big| \, \fu_{i_2} \right) \right] - \kappa \,.\,
\end{align}
% \begin{align}
%     \Pr(\frac{1}{n} \sum_{t=1}^n Y_t > P |  \fu_i) & = \Pr(\frac{1}{n} \sum_{t=1}^n Y_t - (\lambda + \frac{1}{n} \sum_{t=1}^n u_{i,t}) > P - (\lambda + \frac{1}{n} \sum_{t=1}^n u_{i,t}) |  \fu_i)
%     \nonumber\\
%     & = \Pr(\frac{1}{n} \sum_{t=1}^n Y_t - (\lambda + \frac{1}{n} \sum_{t=1}^n u_{i,t}) > A + \delta - \frac{1}{n} \sum_{t=1}^n u_{i,t} |  \fu_i)
%     \nonumber\\
%     & \leq \frac{\text{var}(\frac{1}{n} \sum_{t=1}^n Y_t)}{(A + \delta - \frac{1}{n} \sum_{t=1}^n u_{i,t})^2}
%     \nonumber\\
%     & \leq \frac{\frac{1}{n^2}(n\lambda + \sum_{t=1}^n u_{i,t})}{(A + \delta - \frac{1}{n} \sum_{t=1}^n u_{i,t})^2}
%     \nonumber\\
%     & \leq  \frac{\frac{1}{n^2}(n\lambda + nA)}{\delta^2}
%     \nonumber\\
%     & \leq \frac{\lambda + A}{n\delta^2}
% \end{align}
%%%
    Now, let us focus on the summand in the square brackets in (\ref{Eq.Error_Sum_2}). By (\ref{Eq.Poisson_Channel_Law}), we have
    % ---
    % \begin{align}
    %      W^n(\fy|\fu_{i_1}) - W^n(\fy|\fu_{i_2}) & =
    %      \prod_{t=1}^n \frac{e^{-u_{i_1,t}}u_{i_1,t}^{y_t}}{y_t!} - \prod_{t=1}^n \frac{e^{-u_{i_2,t}}u_{i_2,t}^{y_t}}{y_t!} 
    %     \nonumber\\
    %     & \leq \sum_{t=1}^n \left| \frac{e^{-u_{i_1,t}}u_{i_1,t}^{y_t}}{y_t!} - \frac{e^{-u_{i_2,t}}u_{i_2,t}^{y_t}}{y_t!} \right|
    %     \nonumber\\
    %     & = \sum_{t=1}^n \frac{e^{-u_{i_1,t}}u_{i_1,t}^{y_t}}{y_t!} \left| 1 - e^{-(u_{i_2,t}-u_{i_1,t})} \left(\frac{u_{i_2,t}}{u_{i_1,t}}\right)^{y_t} \right|
    %     \nonumber\\
    %     & = \sum_{t=1}^n \frac{e^{-u_{i_1,t}}u_{i_1,t}^{y_t}}{y_t!} \left| 1 - e^{- \epsilon_n' u_{i_1,t}} \left(\frac{u_{i_2,t}}{u_{i_1,t}}\right)^{y_t} \right|
    %     \nonumber\\
    %     & = \sum_{t=1}^n \frac{e^{-u_{i_1,t}}u_{i_1,t}^{y_t}}{y_t!} \left| 1 - e^{- \epsilon_n' u_{i_1,t}} \left(\frac{u_{i_2,t}}{u_{i_1,t}}\right)^{y_t} \right|
    % \end{align}
    % ---
    \begin{align}
        \label{Ineq.Cond_Channel_Diff}
         W^n \left( \fy \, \big| \, \fu_{i_1} \right) - W^n \left( \fy \, \big| \, \fu_{i_2} \right) 
        & = W^n \left( \fy \, \big| \, \fu_{i_1} \right) \left[1- {W^n \left( \fy \, \big| \, \fu_{i_2} \right)}\,/\,{W^n \left( \fy \, \big| \, \fu_{i_1} \right)}\right] \nonumber\\
        & = W^n \left( \fy \, \big| \, \fu_{i_1} \right) \left[1- \prod_{t=1}^n e^{-(v_{i_2,t}-v_{i_1,t})} \left(\frac{v_{i_2,t}}{v_{i_1,t}}\right)^{y_t} \right] 
        \nonumber\\
        & = W^n \left( \fy \, \big| \, \fu_{i_1} \right) \left[1- \prod_{t=1}^n e^{-\epsilon'_n v_{i_1,t}} \left(1-\epsilon'_n\right)^{y_t} \right], 
        % \nonumber\\
        % &=
        %  \prod_{t=1}^n \frac{e^{-v_{i_1,t}}v_{i_1,t}^{y_t}}{y_t!} - \prod_{t=1}^n \frac{e^{-v_{i_2,t}}v_{i_2,t}^{y_t}}{y_t!} 
        % \nonumber\\
        % & = e^{-\sum_{t=1}^n v_{i_1,t}} \left[  \prod_{t=1}^n \frac{v_{i_1,t}^{y_t}}{y_t!} - e^{-\sum_{t=1}^n (v_{i_2,t} - v_{i_1,t})} \prod_{t=1}^n \frac{v_{i_2,t}^{y_t}}{y_t!} \right] \,.
    \end{align}
    % ---
    where for the last inequality, we employed
    % ---
    \begin{align}
        v_{i_2,t} - v_{i_1,t} \leq \left| v_{i_2,t} - v_{i_1,t} \right| \leq \epsilon'_n v_{i_1,t} \,,\,    
    \end{align}
    % ---
    and
    % ---
    \begin{align}
        1 - \frac{v_{i_2,t}}{v_{i_1,t}}\leq \left| 1 - \frac{v_{i_2,t}}{v_{i_1,t}} \right| \leq \epsilon'_n \,,\,
    \end{align}
    % ---
    which follow from \eqref{Ineq.Converse_Lem_Complement}. 
    Now, we bound the product term inside the bracket as follows:
    % ---
       \begin{align}
       \prod_{t=1}^n e^{-\epsilon'_n v_{i_1,t}}\left(1-\epsilon'_n\right)^{y_t} 
       &=
            e^{- \epsilon'_n\sum_{t=1}^n v_{i_1,t}} \cdot \left( 1 - \epsilon'_n \right)^{\sum_{t=1}^n y_t} \nonumber\\
            & \overset{(a)}{\geq} %e^{-\epsilon'_n\sum_{t=1}^n v_{i_1,t}} \cdot (1-\epsilon'_n)^{n(\lambda + P_{\,\text{max}}+\delta)}
            %\nonumber\\
            %& \geq e^{-n\epsilon'_n(\lambda + P_{\,\text{max}})} \cdot (1-\epsilon'_n)^{nK}
            %\nonumber\\
            %& = 
            e^{-n\epsilon'_n \left( \rho P_{\,\text{max}} + \lambda \right)} \cdot \left( 1 - \epsilon'_n \right)^{n \left( \rho P_{\,\text{max}} + \lambda + \delta \right)}
            \nonumber\\
            & = e^{n\epsilon'_n\delta} \cdot e^{-n\epsilon'_n \left( \rho P_{\,\text{max}} + \lambda + \delta \right)} \cdot \left( 1 - \epsilon'_n \right)^{n \left( \rho P_{\,\text{max}} + \lambda  + \delta \right)}
            \nonumber\\
            & \overset{(b)}{\geq} e^{n\epsilon'_n\delta} \cdot e^{-n\epsilon'_n \left( \rho P_{\,\text{max}} + \lambda + \delta \right)} \cdot \left( 1 - n\epsilon'_n \right)^{ \rho P_{\,\text{max}} + \lambda  + \delta }
            %\nonumber\\
            %& \stackrel{(a)}{\geq} e^{n\epsilon'_n\delta} \cdot e^{-n\epsilon'_n (\lambda + P_{\,\text{max}}+\delta)} \cdot (1-n\epsilon'_n)^{(\lambda + P_{\,\text{max}}+\delta)}
            \nonumber\\
            &\geq e^{n\epsilon'_n \delta} \cdot f(n\epsilon'_n)
            \nonumber\\
            &\stackrel{(c)}{>} f(n\epsilon'_n) 
            \nonumber\\
            &\stackrel{(d)}{\geq} 1 - 3 \left( \rho P_{\,\text{max}} + \lambda + \delta \right) n\epsilon'_n
            \nonumber\\
            & = 1 - \frac{3 \left( \rho P_{\,\text{max}} + \lambda + \delta \right)P_{\,\max}}{n^b}
            \nonumber\\&
             \geq 1 - \kappa \,.\,
            % & \geq e^{-\epsilon'_n n(\lambda+A)} \cdot e^{\frac{nP\epsilon'_n}{\epsilon'_n - 1}}
            % \nonumber\\
            % & = e^{-\frac{\lambda+A}{n^b}} \cdot e^{\frac{nP}{1- n^{1+b}}}
            % \nonumber\\
            % & = e^{\frac{-P-\lambda-A}{n^b}}
            % \nonumber\\
            % & = e^{\frac{-\delta}{n^b}} \geq 1 - \kappa
            % \nonumber\\
            % & = e^{n\epsilon'_n \delta} - e^{-n\epsilon'_n(\lambda+A)}
            % & = \nonumber\\
            % & = e^{\frac{\delta}{n^b}} - e^{-\frac{\lambda+A}{n^b}}
            % \nonumber\\
            % \frac{\epsilon'_n}{\epsilon'_n -1} = \frac{\frac{1}{n^{1+b}}}{\frac{1}{n^{1+b}}-1} = \frac{1}{1- n^{1+b}} \sim \frac{-1}{n^{1+b}}
            \label{Ineq.Cond_Channel_Diff3}
    \end{align}
    % ---
    for sufficiently large $n$. For inequality $(a)$, we used
    % ---
    \begin{align}
        v_{i_1,t} \leq \rho P_{\,\text{max}}+\lambda \,,\, \quad \forall \, t\in[\![n]\!] \,,\,   
    \end{align}
    % ---
    and
    % ---
    \begin{align}
        \sum_{t=1}^n y_t \leq n \left( \rho P_{\max} + \lambda + \delta \right) \,,\,   
    \end{align}
    % ---
    where the latter inequality follows from $\fy\in\B_{i_1}$, cf. \eqref{Eq:Bi1}. For $(b)$, we used Bernoulli's inequality
    % ---
    \begin{align}
        (1 - x)^r \geq 1 - rx \,,\, \quad \forall x > -1 \text{, } \forall r > 0 \,,\,   
    \end{align}
    % ---
    \cite[see Ch.~3]{Mitrinovic13}. For $(c)$, we exploited $e^{n\epsilon'_n \delta}> 1$ and the following definition:
    % ---
    \begin{align}
        f(x) = e^{-cx}(1-x)^c \;,\,
    \end{align}
    % ---
    with $c = \lambda + \rho P_{\,\text{max}}+\delta$. Finally, for $(d)$, we used the Taylor expansion $f(x) = 1-2cx + \mathcal{O}(x^2)$ to obtain the upper bound $f(x) \geq 1-3cx$ for sufficiently small values of $x$.

    % Then, the bound in  (\ref{Ineq.Cond_Channel_Diff3}) becomes
    % % ---
    % \begin{align}
    %     e^{-\epsilon'_n\sum_{t=1}^n v_{i_1,t}} \cdot \left( 1 - \epsilon'_n \right)^{\sum_{t=1}^n y_t} & \geq 1 - 3 \left( \rho P_{\,\text{max}} + \lambda + \delta \right) n\epsilon'_n
    %     \nonumber\\
    %     & = 1 - \frac{3 \left( \rho P_{\,\text{max}} + \lambda + \delta \right)}{n^b}
    %     \nonumber\\
    % & \geq 1 - \kappa \,,\,
    %     \label{Ineq.Cond_Channel_Diff4}
    % \end{align}
    % % ---
    % for sufficiently large $n$, where the equality holds by (\ref{Eq.epsilonn_p}). The last bound implies (\ref{Ineq.Cond_Channel_Diff2}).

    Equation~(\ref{Ineq.Cond_Channel_Diff}) can then be written as follows
    % ---
    \begin{align}
         \label{Ineq.Cond_Channel_Diff2}
         W^n \left( \fy \, \big| \, \fu_{i_1} \right) - W^n \left( \fy \, \big| \, \fu_{i_2} \right) & \leq %W^n(\fy|\fu_{i_1}) - e^{-\sum_{t=1}^n (1+\epsilon'_n) v_{i_1,t}} \cdot (1-\epsilon'_n)^{\sum_{t=1}^n y_t} \cdot \prod_{t=1}^n \frac{v_{i_1,t}^{y_t}}{y_t!}
        % % \nonumber\\
        % % & =  W^n(\fy|\fv_{i_1}) - e^{-2\epsilon'_n\sum_{t=1}^n v_{i_1,t}} \cdot e^{-\sum_{t=1}^n (1-\epsilon'_n) v_{i_1,t}} \cdot (1-\epsilon'_n)^{\sum_{t=1}^n y_t} \cdot \prod_{t=1}^n \frac{v_{i_1,t}^{y_t}}{y_t!} 
        % \nonumber\\
        % & = W^n(\fy|\fu_{i_1}) - e^{-\epsilon'_n\sum_{t=1}^n v_{i_1,t}} \cdot (1-\epsilon'_n)^{\sum_{t=1}^n y_t} \cdot \prod_{t=1}^n \frac{e^{-v_{i_1,t}} v_{i_1,t}^{y_t}}{y_t!} 
        % \nonumber\\
        % & \stackrel{(b)}{=} W^n(\fy|\fu_{i_1}) -  e^{-\epsilon'_n\sum_{t=1}^n v_{i_1,t}} \cdot (1-\epsilon'_n)^{\sum_{t=1}^n y_t} \cdot W^n(\fy|\fu_{i_1}) %
         W^n \left( \fy \, \big| \, \fu_{i_1} \right) \cdot \left[1 - e^{-\epsilon'_n \sum_{t=1}^n v_{i_1,t}} \cdot \left( 1 - \epsilon'_n \right)^{\sum_{t=1}^n y_t} \right] 
        % \nonumber\\
        % & \leq  1 - e^{-\epsilon'_n\sum_{t=1}^n v_{i_1,t}} \cdot (1-\epsilon'_n)^{\sum_{t=1}^n y_t}  \cdot W^n(\fy|\fv_{i_1})
        \nonumber\\
        & \leq \kappa \cdot W^n \left( \fy \, \big| \, \fu_{i_1} \right) \,.
    \end{align}
    % ---
    Combining, (\ref{Eq.Error_Sum_2}), (\ref{Ineq.Cond_Channel_Diff}), and (\ref{Ineq.Cond_Channel_Diff2}) yields
    % ---
    \begin{align}
        P_{e,1}(i_1) + P_{e,2}(i_2,i_1)
        & \overset{(a)}{\geq} 1 - \sum_{\fy\in\B_{i_1}} \left[ W^n \left( \fy \, \big| \, \fu_{i_1} \right) - W^n \left( \fy \, \big| \, \fu_{i_2} \right) \right] - \kappa \nonumber \\ & = 1 - \sum_{\fy \in \B_{i_1}} \left[ \kappa \cdot W^n \left( \fy \, \big| \, \fu_{i_1} \right) \right] - \kappa
        \nonumber\\&
         \overset{(b)}{\geq} 1 - 2\kappa \;,\,
    \end{align}
    % ----
    where for $(a)$, we replaced $\fy\in\B_{i_1}\cap\D_{i_1}$ by $\fy\in\B_{i_1}$ to enlarge the domain and for $(b)$, we used $\sum_{\fy \in \B_{i_1}} W^n \left( \fy \, \big| \, \fu_{i_1}\right)\leq 1$. Clearly, this is a contradiction since the error probabilities tend to zero as $n\rightarrow\infty$. Thus, the assumption in (\ref{Ineq.Converse_Lem_Complement}) is false. This completes the proof of Lemma~\ref{Lem.DConverseFast}.
% -----
\bibliographystyle{IEEEtran}
\bibliography{IEEEabrv,confs-jrnls,Lit}
% ---
\end{document}